\documentclass[aps,pra,twocolumn,superscriptaddress]{revtex4}
\usepackage{bm}
\usepackage{graphicx}
\usepackage{amssymb,amsmath,amsbsy,amsgen,amsfonts}     
\usepackage{dcolumn}
\usepackage{amsthm}
\usepackage{mathrsfs}
\usepackage{latexsym}
\usepackage{array}
\usepackage{amstext}
\usepackage{epsfig}
\usepackage{epstopdf} % for TeXShop on Mark's Mac

\usepackage{color}
\usepackage{units}

\newcommand{\be}{\begin{equation}}
\newcommand{\ee}{\end{equation}}
\newcommand{\ba}{\begin{array}}
\newcommand{\ea}{\end{array}}
\newcommand{\bqa}{\begin{eqnarray}}
\newcommand{\eqa}{\end{eqnarray}}

\begin{document}

\title{Robust entanglement with 3D nonreciprocal photonic topological insulators}

\author{S. Ali Hassani Gangaraj} 
\email{ali.gangaraj@gmail.com}
\address{Department of Electrical Engineering, University of Wisconsin-Milwaukee, 3200 N. Cramer St., Milwaukee, Wisconsin 53211, USA}

\author{George W. Hanson} 
\email{george@uwm.edu}
\address{Department of Electrical Engineering, University of Wisconsin-Milwaukee, 3200 N. Cramer St., Milwaukee, Wisconsin 53211, USA}

\author{Mauro Antezza} 
\email{mauro.antezza@umontpellier.fr}
\address{Laboratoire Charles Coulomb (L2C), UMR 5221 CNRS-Universit de Montpellier, F-34095 Montpellier, France}
\address{Institut Universitaire de France, 1 rue Descartes, F-75231 Paris Cedex 05, France}

\date{\today}

\begin{abstract}
We investigate spontaneous and pumped entanglement of two level systems in the vicinity of a photonic topological insulator interface, which supports a nonreciprocal (unidirectional), scattering-immune and topologically-protected surface plasmon polariton in the bandgap of the bulk material. To this end, we derive a master equation for qubit interactions in a general three-dimensional, nonreciprocal, inhomogeneous and lossy environment. The environment is represented exactly, via the photonic Green function. The resulting entanglement is shown to be extremely robust to defects occurring in the material system, such that strong entanglement is maintained even if the interface exhibits electrically-large and geometrically sharp discontinuities. Alternatively, depending on the initial excitation state, using a non-reciprocal environment allows two qubits to remain unentangled even for very close spacing. The topological nature of the material is manifest in the insensitivity of the entanglement to variations in the material parameters that preserve the gap Chern number. Our formulation and results should be useful for both fundamental investigations of quantum dynamics in nonreciprocal environments, and technological applications related to entanglement in two-level systems.
\end{abstract}

\maketitle

%%%%%%%%%%%%%%%%%%%%%%%%%%%%%%%%%%%%%%%%%%%%%%
%%%%%%%%%%%%%%%%%%%%%%%%%%%%%%%%%%%%%%%%%%%%%%
%%%%%%%%%%%%%%%%%%%%%%%%%%%%%%%%%%%%%%%%%%%%%%

\section{Introduction}

Entanglement as a quantum resource is important for a range of emerging applications, including quantum computing \cite{5} and quantum cryptography \cite{6}. A main obstacle to the development of entanglement-based systems is decoherence associated with the unavoidable coupling between a quantum system and the degrees of freedom of the surrounding environment \cite{10}. However, reservoir engineering methods have changed the idea of trying to minimize coupling to the environment to one of modifying the properties of the environment in order to achieve a desired state. These methods include using dissipative dynamics \cite{Plenio02,Hartmann,Verstraete,Krauter,Lucas,Sarlette}, recently extended to systems out of thermal equilibrium \cite{Bellomo12,Bellomo13,Bellomo13EPL,Bellomo13NJP,Bellomo15}, as well as, e.g., exploiting the effect of measurements and feedback to achieve a desired final state \cite{Mancini,Stevenson}. 

Another emerging resource for reservoir engineering is the use of nonreciprocal environments \cite{Zoller3}. In particular, there has been considerable investigation of quantum spin networks in chiral waveguides \cite{Garcia2,Zoller0,Zoller1,Zoller2,MLS,SJ,MS}. The previous work on spin dynamics in quantum chiral environments has focused on one-dimensional (1D) waveguide models. Here, we investigate two-level (spin) quibit interactions mediated by uni-directional surface-plasmon-polaritions (SPPs) at the interface of a photonic topological insulator (PTI) and a topological-trivial material. 

PTIs represent a broad class of materials that are attracting wide interest for both fundamental and applied reasons \cite{Haldane,Rechtsman,Lu,Kh}. Perhaps their most celebrated aspect is their ability to support SPPs that are unidirectional, propagate in the bulk bandgap, and are topologically protected from backscattering at discontinuities \cite{Joannopoulos,Veronis,Joannopoulos2,Rechtsman1,Arthur,Mario2,Hassani2,Hassani3}. PTIs can be broadly divided into two classes, (i) those with broken time reversal (TR) symmetry, which are photonic analogs of quantum Hall insulators (photonic quantum Hall effect (PQHE)), and (ii) those that are time-reversal-invariant but have broken inversion symmetry, which are photonic analogs of electronic topological insulators/quantum spin Hall insulators (photonic quantum spin Hall Effect (PQSHE)). Although as a specific example we consider PTIs of the PQHE type, the formulation presented here is general. 

In this work, we develop a master equation (ME) for three-dimensional (3D), nonreciprocal, inhomogeneous and lossy environments, based on the macroscopic canonical quantization scheme described in \cite{Welsch0,Welsch1, Trung}, extended to nonreciprocal media \cite{Buh}. In Section \ref{ME1} we derive the master equation, and in Section \ref{CS} we present the equations for concurrence as a measure of entanglement. In Section \ref{NR} we consider the topological aspect of concurrence for a PQHE-type PTI system consisting of a plasma continuum. Then, qubit entanglement dynamics are examined for several waveguiding systems. We focus of the aspects unique to the topological and nonreciprocal environment, such as the preservation of entanglement in the presence of large defects. Three appendices present a discussion of various approximations used in the development of the ME, a comparison with previous 1D chiral MEs, and a derivation of the unidirectional concurrence.

%%%%%%%%%%%%%%%%%%%%%%%%%%%%%%%%%%%%%%%%%%%%%%
%%%%%%%%%%%%%%%%%%%%%%%%%%%%%%%%%%%%%%%%%%%%%%
%%%%%%%%%%%%%%%%%%%%%%%%%%%%%%%%%%%%%%%%%%%%%%

\section{Theoretical model}
\begin{figure}[h]
	\begin{center}
		\noindent
		\includegraphics[width=3.5in]{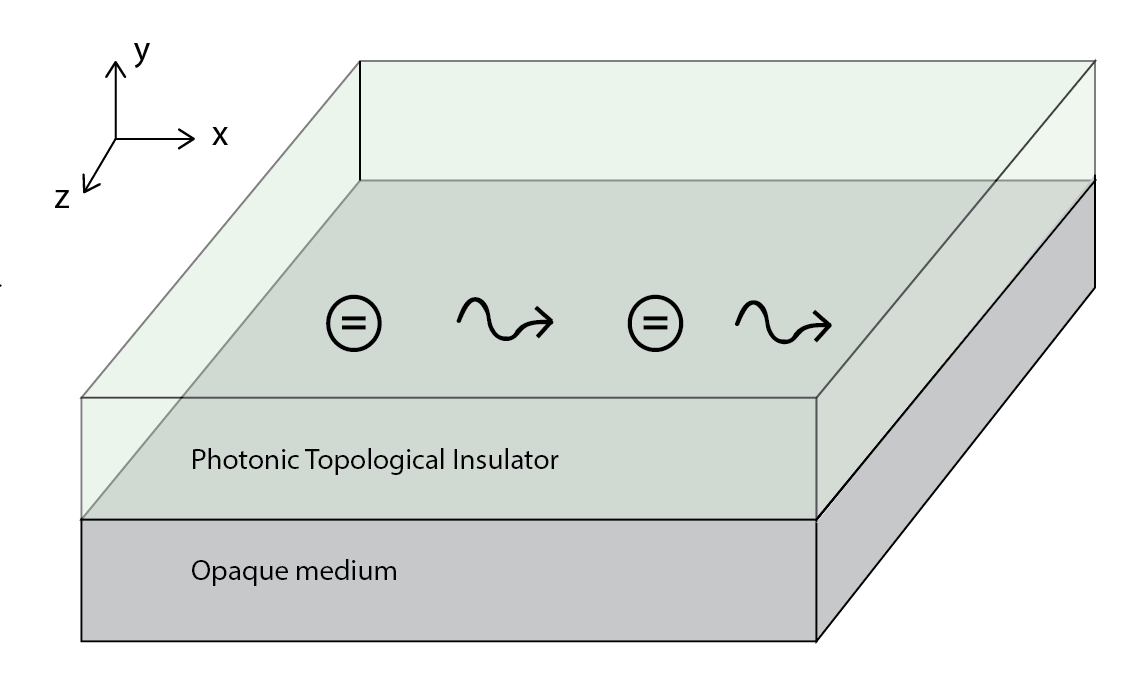}
		\caption{Two qubits at the interface of a PTI and topologically trivial medium. The resulting unidirectional SPP provides a strongly non-reciprocal environment for qubit entanglement.}\label{geom}
	\end{center}
\end{figure}

In this section, we first derive a general ME valid for both reciprocal and nonreciprocal, inhomogeneous and lossy environments. This form is valid for 3D, 2D and 1D systems since it is expressed in terms of the electromagnetic Green function. Then, we present concurrence expressions for the unidirectional case. The physical system we will consider is that of two qubits at the interface of a PTI and another (eventually topologically  trivial) medium, as depicted in Fig. \ref{geom}, although the development is completely general.

\subsection{Master equation for general 3D nonreciprocal environments}\label{ME1}
We consider qubits with transition frequency $ \omega_0 $ interacting through a general nonreciprocal environment. For a derivation in the reciprocal case, see \cite{SteveH}. 

The classical electric field satisfies
\begin{equation}
	\left[\nabla \times \mathbf{\mu}^{-1}(\mathbf{r},\omega) \nabla \times - \frac{\omega^2}{c^2} \mathbf{\varepsilon}(\mathbf{r},\omega) \right] \boldsymbol{\mathrm{E}}(\mathbf{r},\omega) = i \omega \mu_0 \boldsymbol{\mathrm{j}}_s(\mathbf{r},\omega),
\end{equation}
where $c$ is the vacuum speed of light, $ \mu(\mathbf{r},\omega) $ and $ \boldsymbol{\varepsilon}(\mathbf{r},\omega) $ are the material permeability and permittivity, and $ \boldsymbol{\mathrm{j}}_s(\mathbf{r},\omega) $ is the noise current. In this work, we suppose that the medium is non-magnetic, $ \mathbf{\mu} (\mathbf{r},\omega) =\mathbf{I} $, where $\boldsymbol{ \mathrm{I}} $ is the unit dyad, but that the permittivity is a tensorial quantity. By defining the noise current in terms of polarization as $ \boldsymbol{\mathrm{j}}_s = -i \omega \boldsymbol{\mathrm{P}}_s $, which is associated with material absorption by the fluctuation-dissipation theorem, the electric field Green tensor is the solution of  
\begin{equation}
	\left[\nabla \times  \nabla \times - \frac{\omega^2}{c^2} \mathbf{\varepsilon}(\mathbf{r},\omega) \right] \boldsymbol{\mathrm{G}}(\mathbf{r} , \mathbf{r}',\omega) = \boldsymbol{\mathrm{I}} \delta(\mathbf{r} , \mathbf{r}')
\end{equation}
and the electric field is $ \boldsymbol{\mathrm{E}}(\mathbf{r},\omega)  = (\omega^2 / c^2 \varepsilon_0) \int_V d \mathbf{r}' \boldsymbol{\mathrm{G}}(\mathbf{r}, \mathbf{r}', \omega) \cdot \boldsymbol{\mathrm{P}}_s(\mathbf{r}' , \omega)  $. Following the standard macroscopic canonical quantization \cite{Welsch0,Welsch1, Trung}, the noise polarization can be expressed in term of the bosonic field annihilation operator as \cite{Buh}
\begin{equation}
	\hat{\boldsymbol{\mathrm{P}}}_s(\mathbf{r},\omega) = -i \sqrt{ \frac{\hbar \varepsilon_0}{\pi}} \boldsymbol{\mathrm{T}}(\mathbf{r},\omega) \cdot \hat{\boldsymbol{\mathrm{f}}} (\mathbf{r},\omega), 
\end{equation}
where 
\begin{equation}
	\boldsymbol{\mathrm{T}}(\mathbf{r},\omega) \cdot \boldsymbol{\mathrm{T}}^{\dagger}(\mathbf{r},\omega) = \frac{1}{2i} \left[ \boldsymbol{\varepsilon}(\mathbf{r},\omega) - \boldsymbol{\varepsilon}^{\dagger}(\mathbf{r},\omega)  \right],
\end{equation}
and, for the special case of a symmetric permittivity tensor (e.g., a reciprocal medium), $ \boldsymbol{\mathrm{T}}(\mathbf{r},\omega) = \sqrt{ \mathrm{Im} \boldsymbol{\varepsilon} (\mathbf{r},\omega) }   $. The bosonic field operators $ \hat{\boldsymbol{\mathrm{f}}} (\mathbf{r},\omega) $ obey the commutation relations $  [   \hat{\mathrm{f}}_j (\mathbf{r},\omega) , \hat{\mathrm{f}}^{\dagger}_{j'} (\mathbf{r}',\omega')   ] = \delta_{jj'} \delta(\mathbf{r} - \mathbf{r}') \delta(\omega - \omega') $ and $ [   \hat{\mathrm{f}}_j (\mathbf{r},\omega) , \hat{\mathrm{f}}_{j'} (\mathbf{r}',\omega')   ] = 0 $. The noise polarization operator generates the electric field operator 
\begin{equation}\label{Eq:E(G)}
	\hat{\boldsymbol{\mathrm{E}}}(\mathbf{r},\omega) = i \sqrt{\frac{\hbar}{\pi \varepsilon_0}} \frac{\omega^2}{c^2} \int d \mathbf{r}' \boldsymbol{\mathrm{G}}(\mathbf{r}, \mathbf{r}' , \omega) \cdot \boldsymbol{\mathrm{T}}(\mathbf{r},\omega) \cdot \hat{\boldsymbol{\mathrm{f}}}(\mathbf{r}', \omega),
\end{equation}
where $ \boldsymbol{\mathrm{G}}(\mathbf{r}, \mathbf{r}' , \omega)  $ is the classical electric field Green tensor. The nonreciprocal Green tensor has the following useful property \cite{Buh}
\begin{align}
	& 2i\frac{\omega ^{2}}{c^{2}}\int d^{3}\mathrm{r}^{\prime \prime }%
	\boldsymbol{\mathrm{G}}(\mathbf{r},\mathbf{r}^{\prime \prime },\omega )\cdot 
	\boldsymbol{\mathrm{T}}(\mathbf{r}^{\prime \prime },\omega )\cdot 
	\boldsymbol{\mathrm{T}}^{\dagger }(\mathbf{r}^{\prime \prime },\omega )%
	\boldsymbol{\mathrm{G}}^{\dag }(\mathbf{r}^{\prime },\mathbf{r}^{\prime
		\prime },\omega )  \nonumber \\
	& =\boldsymbol{\mathrm{G}}(\mathbf{r},\mathbf{r}^{\prime },\omega )-%
	\boldsymbol{\mathrm{G}}^{\dagger }(\mathbf{r}^{\prime },\mathbf{r},\omega ).
\end{align}

Under the dipole approximation, the governing Hamiltonian of a system of qubits (two level atoms) interacting with the surrounding environment can be written as
\begin{align}
	\mathrm{H} & = \int d^3 \mathbf{r} \int_{0}^{\infty} d \omega \hbar \omega  \hat{\boldsymbol{\mathrm{f}}}^{\dagger} (\mathbf{r},\omega)   \hat{\boldsymbol{\mathrm{f}}} (\mathbf{r},\omega) + \sum_i \hbar \omega_i \hat{\sigma}^{\dagger}_i {\sigma}_i \notag \\ & - \sum_i \int_{0}^{\infty} d \omega ( \hat{\boldsymbol{\mathrm{d}}}_i \cdot \boldsymbol{\mathrm{E}}(\mathbf{r}_i , \omega) + \mathrm{H.c.}  ),
\end{align}
where the right side can be decomposed into the reservoir Hamiltonian $ \mathrm{H}_r $ (first term), the qubit Hamiltonian $ \mathrm{H}_s $ (second term), and the interaction Hamiltonian $ \mathrm{H}_{sr} $ (third term). We can modify the total Hamiltonian to include the coherent drive (external laser pump) Hamiltonian $ \mathrm{V}^{AF} $, given later ((\ref{Eq:laser})). We transform to a frame rotating with the laser frequency $ \omega_l $ ($ \mathrm{H} \rightarrow \hat{\mathrm{U}}^{\dagger}(t) \mathrm{H} \hat{\mathrm{U}}(t),~ \hat{\mathrm{U}}(t) = e^{-i\omega_l \sum_i {\sigma}^{\dagger}_i {\sigma}_i t  }  $) and write the total density matrix of the qubit system and reservoir according to the Schr$\ddot{\mathrm{o}}$dinger equation $ \partial_t \rho_T = -i [\mathrm{H},\rho_T]/\hbar $, then we transform to the interaction picture ($ \hat{\mathrm{O}}_I = \hat{\mathrm{U}}^{\dagger}(t) \mathrm{H} \hat{\mathrm{U}}(t),~ \hat{\mathrm{U}}(t) = e^{-i (\mathrm{H}_s + \mathrm{H}_r)t/\hbar} $) where $ \partial_t \rho_{T,I} = -i [ \mathrm{H}_I, \rho_{T,I} ] $ with $ \mathrm{H}_I = \mathrm{H}_{sr,I} $. We integrate to find
\begin{equation}\label{rho_TI}
	\rho_{T,I} = \rho_{I}(0) \mathrm{R}_0 + \frac{-i}{\hbar} \int_{0}^{t} dt' [\mathrm{H}_I(t'), \rho_{T,I}(t') ]
\end{equation} 
where $ \mathrm{R}_0 $ is the initial reservoir density matrix. In the interaction picture, by considering $ \Gamma_{ii} \ll \omega $ for optical frequencies we make the rotating wave approximation (RWA) in $ \mathrm{H}_I $ and drop the rapidly varying counter-rotating terms proportional to $ {\sigma}^{\dagger}(t') \boldsymbol{\mathrm{f}}^{\dagger}(\mathbf{r}', \omega ) e^{i(\omega_l + \omega)t'} $ and its Hermitian conjugate. The interaction Hamiltonian in the interaction picture reduces to 
\begin{equation}
	\mathrm{H}_I(t) = - \sum_i \left(  \int_{0}^{\infty} d \omega {\sigma}^{\dagger}_i(t) \boldsymbol{\mathrm{d}}_i \cdot \boldsymbol{\mathrm{E}}(\mathbf{r}_i, \omega) e^{-i(\omega - \omega_l)t} + \mathrm{H.c.}  \right)
\end{equation}

To find the system density matrix we insert (\ref{rho_TI}) into the interaction picture Schr$\ddot{\mathrm{o}}$dinger equation and trace over the reservoir,
\begin{align}\label{Eq:29}
	\partial_t \rho_{I} & = \mathrm{Tr}_R  \{  \frac{-i}{\hbar} \left[  \mathrm{H}_I, \rho_{I}(0) \mathrm{R}_{0,I}  \right]    \} \notag \\ & - \frac{1}{\hbar^2} \int_{0}^{t} dt' \mathrm{Tr}_R \{ \left[   \mathrm{H}_I(t), \left[   \mathrm{H}_I(t'), \rho_{T,I}(t') \right] \right] \}.
\end{align}

Aside from the rotating wave approximation, we apply a number of other approximations to the density matrix to simplify this further (see Appendix I). We first take the mean initial system reservoir coupling to be zero such that $ \mathrm{Tr}_R  \{  \frac{-i}{\hbar} \left[  \mathrm{H}_I, \rho_{I}(0) \mathrm{R}_{0,I}  \right]    \} = 0 $. Then we apply the Born approximation, which states that the reservoir will be largely unaffected by its interaction by the system. Next, we assume that the evolution of the density matrix only depends on its current state (Born-Markov approximation). The Born-Markov approximation comes from the assumption that the reservoir relaxation time is much faster than the relaxation time of the system, and so the memory effect of the reservoir can be ignored. Lastly, we make a second Markov approximation, extending the upper limit of the time integral to infinity to produce a fully Markovian equation. With these simplifications we have 
\begin{equation}\label{rho_interaction}
	\partial_t \rho_{I} = - \frac{1}{\hbar^2} \int_{0}^{\infty} dt' \mathrm{Tr}_R \left\{ \left[ \mathrm{H}_I(t), \left[ \mathrm{H}_I(t-t'),\rho_{I}(t) \mathrm{R}_0 \right]  \right] \right\}.
\end{equation}

We suppose that the atomic transition frequency of the qubits is $ \omega_0 $. Then, for the first term in (\ref{rho_interaction}) we have
\begin{align}
	& \mathrm{Tr}_R \left\{ \mathrm{H}_I(t)\mathrm{H}_I(t-t') \rho_{I}(t) \mathrm{R}_0\right\} =   \sum_{i,j} \mathrm{d}_{\alpha i} \mathrm{d}_{\beta j} \notag \\ & \int_{0}^{\infty} d\omega  e^{i(\omega_0 - \omega)t'} {\sigma}^{\dagger}_i {\sigma}_j \rho_{T,I}   \mathrm{Tr}_R \left( \hat{\mathrm{E}}_{\alpha}(\mathbf{r}_i,\omega) \hat{\mathrm{E}}^{\dagger}_{\beta}(\mathbf{r}_j,\omega) \mathrm{R}_0  \right) 
\end{align}
where
\begin{equation}
{\sigma_i} = \left |g_i \right > \left <e_i \right |,~ {\sigma_i}^{\dagger} = \left |e_i \right > \left <g_i \right |
\end{equation}
are the atomic lowering/raising operators describing energy level transitions for each qubit, and where it is supposed that one of the qubits is polarized along $ \alpha $ and the other one is polarized along $ \beta $. Considering (\ref{Eq:E(G)}) for the nonreciprocal Green tensor and $ \mathrm{Tr}_R \{ \hat{\boldsymbol{\mathrm{f}}}(r,\omega) \hat{\boldsymbol{\mathrm{f}}}^{\dagger}(r',\omega') \mathrm{R}_0  \} = (\bar{n}(\omega) + 1) \delta(r-r') \delta(\omega - \omega')$ with zero thermal photon occupation $ \bar{n}(\omega) = 0 $, it can be easily shown that 
\begin{align}
	&\mathrm{Tr}_R \left( \mathrm{E}_{\alpha}(\mathbf{r}_i,\omega) \mathrm{E}^{\dagger}_{\beta}(\mathbf{r}_j,\omega) \mathrm{R}_0  \right) =  \frac{\hbar}{\pi \varepsilon_0} \frac{\omega^4}{c^4} \notag \\ & \int d^3 \mathbf{r} \mathrm{G}_{\alpha \gamma}(\mathbf{r}_i,\mathbf{r},\omega)  \left[ \frac{{\varepsilon}_{\gamma \gamma'}(\mathbf{r},\omega) - {\varepsilon}^{\dagger}_{\gamma \gamma'}(\mathbf{r},\omega)}{2i}  \right]  \mathrm{G}^*_{ \gamma' \beta}(\mathbf{r}_j,\mathbf{r},\omega) \notag \\&
	= \frac{\hbar}{2i \pi \varepsilon_0} \frac{\omega^2}{c^2} \left( \mathrm{G}_{\alpha, \beta }(\mathbf{r}_i,\mathbf{r}_j,\omega) - \mathrm{G}^*_{\beta,\alpha}(\mathbf{r}_j,\mathbf{r}_i,\omega)  \right).
\end{align}
Thus, we have
\begin{align}\label{Eq:2nd}
	&\mathrm{Tr}_R \left\{ \mathrm{H}_I(t)\mathrm{H}_I(t-t') \rho_{I}(t) \mathrm{R}_0  \right\}   = \frac{\hbar}{2i\pi \varepsilon_0 c^2 }   \sum_{i,j} {\sigma}^{\dagger}_i {\sigma}_j \rho_{I}(t) \notag \\& \int_{0}^{\infty}  \left(  \mathrm{d}_{\alpha i} \mathrm{G}_{\alpha \beta}(\mathbf{r}_i,\mathbf{r}_j,\omega) \mathrm{d}_{\beta j} -  \mathrm{d}_{\beta j} \mathrm{G}^*_{\beta \alpha }(\mathbf{r}_j,\mathbf{r}_i,\omega) \mathrm{d}_{\alpha i}  \right)   \notag \\ & \omega^2 d\omega e^{i(\omega_0 - \omega)t'}.
\end{align}
Following the same procedure for the second term in (\ref{rho_interaction}),
\begin{align}\label{Eq:1st}
	& \mathrm{Tr}_R \left\{ \mathrm{H}_I(t-t')\rho_{I}(t) \mathrm{R}_0 \mathrm{H}_I(t)  \right\}   = \frac{\hbar }{2i\pi\varepsilon_0c^2}  \sum_{i,j} {\sigma}_j \rho_{I}(t) {\sigma}^{\dagger}_i \notag \\ &   \int_{0}^{\infty}    \left(\mathrm{d}_{\beta i} \mathrm{G}_{\beta \alpha}(\mathbf{r}_i,\mathbf{r}_j,\omega) \mathrm{d}_{\alpha j} -  \mathrm{d}_{\alpha j} \mathrm{G}^*_{ \alpha \beta }(\mathbf{r}_j,\mathbf{r}_i,\omega) \mathrm{d}_{\beta i}\right)   \notag \\ & \omega^2 d\omega e^{i(\omega_0 - \omega)t'}.
\end{align}
Replacing (\ref{Eq:2nd}) and  (\ref{Eq:1st}) in (\ref{rho_interaction}) and performing the time integral over $ t' $ gives the evolution of the density matrix in the interaction picture, where we have used the Kramers-Kronig relation 
\begin{align}
	&\mathcal{P} \int_{-\infty}^{\infty} \frac{\mathrm{Re} \mathrm{G}_{\alpha \beta} }{\omega - \omega_0} d \omega = - \pi \mathrm{Im} \mathrm{G}_{\alpha \beta} \notag \\&
	\mathcal{P} \int_{-\infty}^{\infty} \frac{\mathrm{Im} \mathrm{G}_{\alpha \beta} }{\omega - \omega_0} d \omega = \pi \mathrm{Re} \mathrm{G}_{\alpha \beta}.
\end{align}
Transforming back to the Schr$\ddot{\mathrm{o}}$dinger picture, we obtain the master equation for the two-level system dynamics 
\begin{align}\label{Eq:ME}
\partial_t \rho_s(t) = -\frac{i}{\hbar} \left[ \mathrm{H}_s +\mathrm{V}^{AF}, {\rho_s}(t) \right] +  \mathcal{L} {\rho}(t), 
\end{align}
where
\begin{align}\label{Eq:Lindblad_bi}
\mathcal{L} {\rho}_{s}(t) & = \notag \\&
\sum_{i}\frac{\Gamma _{ii}(\omega _{0})
}{2}\left( 2{\sigma}_{i} {\rho}_{s}(t){\sigma}_{i}^{\dagger }-
{\sigma}_{i}^{\dagger }{\sigma}_{i} {\rho}_{s}(t)- {\rho}
_{s}(t){\sigma}_{i}^{\dagger }{\sigma}_{i}\right) \notag \\&
+ \sum^{i \neq j}_{i,j} \frac{\Gamma_{ij} (\omega_0) }{2} \left(  \left[ {\sigma}_j {\rho_s}(t), {\sigma}^{\dagger}_i  \right] +  \left[ {\sigma}_i , {\rho_s}(t) {\sigma}^{\dagger}_j  \right]  \right) \notag \\&
+ \sum^{i \neq j}_{i,j} g_{ij} (\omega_0)  \left(  \left[ {\sigma}_j {\rho_s}(t), -i{\sigma}^{\dagger}_i  \right] +  \left[ i{\sigma}_i , {\rho_s}(t) {\sigma}^{\dagger}_j  \right]  \right).
\end{align}

Equation (\ref{Eq:Lindblad_bi}) is one of the primary results of this work, and is applicable to both reciprocal and nonreciprocal environments and an arbitrary number of qubits. In (\ref{Eq:Lindblad_bi}), $\mathcal{L}$ is the Lindblad superoperator for the general nonreciprocal medium, involving the dissipative decay rate, $ \Gamma_{ij}(\omega_0) $, and the coherent coupling terms, $ g_{ij}(\omega_0) $, in terms of the electromagnetic Green dyadic,
\begin{align}\label{coupling_terms}
& \Gamma _{ij}(\omega _{0}) =\frac{2\omega _{0}^{2}}{\varepsilon _{0}\hbar
	c^{2}}\sum_{\alpha ,\beta =x,y,z}\mathrm{d}_{\alpha i}\mathrm{Im}\left(\mathrm{G}_{\alpha \beta }(%
\mathbf{r}_{i},\mathbf{r}_{j},\omega _{0})\right)\mathrm{d}_{\beta j}, \notag \\&
g_{ij}(\omega _{0})=\frac{\omega _{0}^{2}}{\varepsilon _{0} \hbar c^{2}}
\sum_{\alpha ,\beta =x,y,z} \mathrm{d}_{\alpha i}  \mathrm{Re}(\mathrm{G}_{\alpha\beta}(\mathbf{r}_{i},\mathbf{r}
_{j},\omega )) \mathrm{d}_{\beta j}.
\end{align}
The Hamiltonian of the decoupled qubits is 
\begin{equation}
\mathrm{H}_s = \sum_{i} \hbar \Delta \omega_i {\sigma}^{\dagger}_i {%
	\sigma}_i,
\end{equation}
where $\Delta \omega _{i}=\omega _{0}-\omega_l-\delta _{i} $, with $\delta
_{i}=g_{ii}$ being the Lamb shift and $\omega_l $ is the laser frequency of an external source. The Lamb shift for optical emitters is in general on the order of a few GHz, therefore the effect of the Lamb-shift for optical frequencies is small $(\omega _{i}\sim 10^{15}~\mathrm{Hz},~\delta _{i}\sim 10^{9}~\mathrm{Hz})$, and can be ignored, or assumed to be accounted for in the definition of the transition frequency $\omega_0$. In (\ref{Eq:ME}), the term
\begin{align}\label{Eq:laser}
\mathrm{V}^{AF} & = - \hbar \left( \Omega_1  e^{- i \Delta_l t} \sigma_1^{\dagger} + \Omega_1^* e^{i \Delta_l t} \sigma_1 \right) \notag \\ & 
~~~-\hbar \left( \Omega_2  e^{- i \Delta_l t} \sigma_2^{\dagger} + \Omega_2^* e^{i \Delta_l t} \sigma_2 \right)
\end{align}
represents the external coherent drive applied to each qubit at laser frequency $ \omega_l $. Due to its large amplitude we treat the drive field as a $c$-number where $\Omega_i = \mathbf{d}_{i} \cdot \mathbf{E}_{0}^i / \hbar$ is a Rabi frequency and $\Delta_l = \omega_0 - \omega_l$ is the detuning parameter. 

For the reciprocal case where $ \Gamma_{ij} = \Gamma_{ji} $ and $ g_{ij} = g_{ji} $ it can be shown that (\ref{Eq:Lindblad_bi}) is the well-known reciprocal (bidirectional) master equation \cite{ME_O, Garcia}. In the reciprocal case, some terms associated with $g_{ij}=g_{ji}$ cancel each other out and are eliminated from the dissipative term. For example, ${\sigma}_{i} {\rho}_{s}(t){\sigma}_{i}^{\dagger}, i \neq j$, appears in the nonreciprocal case but is absent in the reciprocal case.

For a system of two qubits, \eqref{Eq:Lindblad_bi} can be written in the simple form  
\begin{align}\label{OME}
	&\mathcal{L} {\rho}_{s}(t) = \sum_{j=1,2} \frac{\Gamma_{jj}}{2} \left(2 \sigma_j \rho_s \sigma^{\dagger}_j - \rho_s \sigma^{\dagger}_j \sigma_j - \sigma^{\dagger}_j \sigma_j \rho_s  \right) \notag \\ & 
	+ \left( \frac{\Gamma_{21}}{2} + i g_{21}  \right) \left( \sigma_2 \rho_s \sigma^{\dagger}_1 - \rho_s \sigma^{\dagger}_1 \sigma_2  \right) \notag \\ & + \left( \frac{\Gamma_{21}}{2} - i g_{21}  \right) \left( \sigma_1 \rho_s \sigma^{\dagger}_2 - \sigma^{\dagger}_2 \sigma_1 \rho_s \right) \notag \\ & 
	+ \left( \frac{\Gamma_{12}}{2} + i g_{12}  \right) \left( \sigma_1 \rho_s \sigma^{\dagger}_2 - \rho_s \sigma^{\dagger}_2 \sigma_1  \right) \notag \\ & + \left( \frac{\Gamma_{12}}{2} - i g_{12}  \right) \left( \sigma_2 \rho_s \sigma^{\dagger}_1 - \sigma^{\dagger}_1 \sigma_2 \rho_s \right).
	\end{align}
A comparison with previous 1D chiral ME formulations is provided in Appendix II.
%%%%%%%%%%%%%%%%%%%%%%%%%%%%%%%%%%%%%%%%%%%%%%
%%%%%%%%%%%%%%%%%%%%%%%%%%%%%%%%%%%%%%%%%%%%%%

\subsection{Transient entanglement: Unidirectional SPP-assisted qubit communication}\label{CS}
In this work, all numerical results are computed using the master equation (\ref{Eq:ME}) with the general 3D Lindblad superoperator (\ref{Eq:Lindblad_bi}), where the Green tensor for complicated environments is obtained numerically. However, as shown in Appendix III, if the system of qubits are communicating through a strongly nonreciprocal environment e.g., $ {\boldsymbol{\mathrm{G}}}({\boldsymbol{\mathrm{r}}}_1,{\boldsymbol{\mathrm{r}}}_2) = 0 $ ($ \Gamma_{12}=g_{12}=0 $) and $ {\boldsymbol{\mathrm{G}}}({\boldsymbol{\mathrm{r}}}_2,{\boldsymbol{\mathrm{r}}}_1) \neq 0 $, then the concurrence (as a measure of entanglement \cite{Wootters}) is  
\begin{align}\label{Eq:Ct}
\mathcal{C}(t) & =  2 \sqrt{ \frac{  \Gamma^2_{21}}{4} +  g^2_{21} }t e^{-\Gamma_{11}t} \notag \\ &
= 2 \frac{\omega^2_0 \mathrm{d}_y^2}{\hbar \varepsilon_0 c^2} |\boldsymbol{\mathrm{G}}_{yy}(\boldsymbol{\mathrm{r}}_2, \boldsymbol{\mathrm{r}}_1, \omega_0 )| t e^{-\Gamma_{11}t},
\end{align}
where it has been assumed that the qubits are both polarized along the $y$-axis. This is the general unidirectional result. Concurrence reaches its maximum value at $ t = 1/\Gamma_{11} $, such that $ \mathcal{C}_{\mathrm{max}}= 2 \sqrt{\tilde{\Gamma}^2_{12}/4 + \tilde{g}^2_{21} }/e$, where $ \tilde{\Gamma}_{21} $ and $ \tilde{g}_{21} $ are rates normalized by $ \Gamma_{11} $.

Although the Hamiltonian in nonreciprocal systems is non-Hermitian, it can be seen (Appendix III) that the density matrix is Hermitian, probability conservation holds ($\mathrm{Tr}(\rho) =1$), and that diagonal elements of the density operator can be interpreted as population densities, as for Hermitian Hamiltonians.

For two identical qubits interacting through a reciprocal medium, 
\begin{align}
\mathcal{C}_{\mathrm{recip}}(t)& =\left\{ \frac{1}{4}\left[ e^{-(\Gamma
_{11}+\Gamma _{12})t}-e^{-(\Gamma _{11}-\Gamma _{12})t}\right] ^{2}\right.  
\nonumber  \label{Eq:Ct_recip} \\
& \left. +e^{-2\Gamma _{11}t}\sin ^{2}(2g_{12}t)\right\} ^{1/2}.
\end{align}

One of the main differences between the concurrence in the reciprocal case (\ref{Eq:Ct_recip}), and in the unidirectional case (\ref{Eq:Ct}), is the presence of the sinusoidal term in (\ref{Eq:Ct_recip}). When $ g_{12} $ is strong enough, this sinusoidal term causes oscillations in the transient concurrence related to photons being recycled between the two qubits, with a period that corresponds to the round trip time of the coupled qubits through the reciprocal medium (Rabi oscillations). For the unidirectional case (\ref{Eq:Ct}) Rabi oscillations can not occur.

It was shown in \cite{Garcia} that for qubits coupled to an infinite reciprocal waveguide system, the positions of $ \Gamma_{ij} $ maxima/minima correspond to positions of $ g_{ij} $ minima/maxima (for finite waveguides, see \cite{H1}). Thus, in general, coherent and dissipative regimes become dominant at different separations between emitters. It was further shown in \cite{Garcia} that for an infinite reciprocal plasmonic waveguide the best entanglement was obtained when $ \Gamma_{ij} $ was large and $ g_{ij} $ was small (forming the dissipative regime), which forces a restriction on the positioning of the qubits in the reciprocal case. However, in the unidirectional case the qubit positioning is unimportant, as detailed in \cite{Garcia2}, and the qubits can be anywhere in the coherent or dissipative regimes, which is a practical advantage of these unidirectional systems. 

\begin{figure}[h]
	\begin{center}
		\noindent
		\includegraphics[width=3.5in]{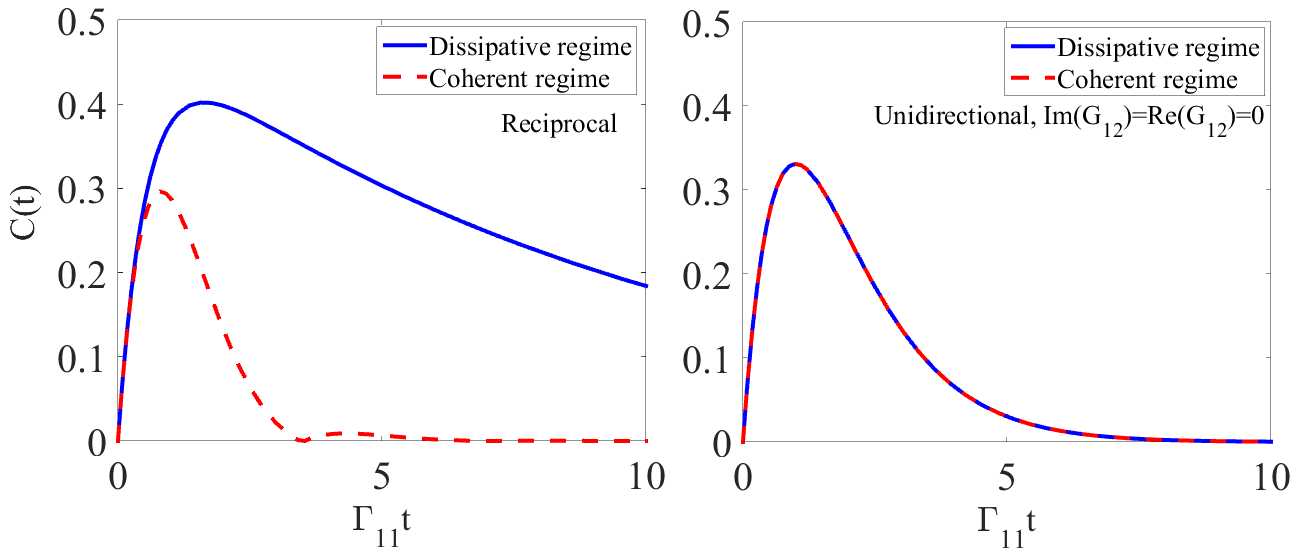}
		\caption{Left panel: Concurrence between two qubits for a reciprocal system. For the dissipative regime $ \mathrm{Im}(\mathrm{G}(\boldsymbol{\mathrm{r}}_1, \boldsymbol{\mathrm{r}}_2)) = \mathrm{Im}(\mathrm{G}(\boldsymbol{\mathrm{r}}_2, \boldsymbol{\mathrm{r}}_1)) = 0.9 $ and $  \mathrm{Re}(\mathrm{G}(\boldsymbol{\mathrm{r}}_1, \boldsymbol{\mathrm{r}}_2))=0  $ and for the coherent regime $ \mathrm{Re}(\mathrm{G}(\boldsymbol{\mathrm{r}}_1, \boldsymbol{\mathrm{r}}_2)) = \mathrm{Re}(\mathrm{G}(\boldsymbol{\mathrm{r}}_2, \boldsymbol{\mathrm{r}}_1)) = 0.9 $ and $  \mathrm{Im}(\mathrm{G}(\boldsymbol{\mathrm{r}}_1, \boldsymbol{\mathrm{r}}_2)) =0 $. Right panel: Concurrence between two qubits for a unidirectional system. For the dissipative regime $ \mathrm{Im}( \boldsymbol{\mathrm{G}}( \boldsymbol{\mathrm{r}}_2 , \boldsymbol{\mathrm{r}}_1 )  ) =  0.9, ~ \mathrm{Re}( \boldsymbol{\mathrm{G}}( \boldsymbol{\mathrm{r}}_2 , \boldsymbol{\mathrm{r}}_1 )  ) = 0 $ and for the coherent case $ \mathrm{Im}( \boldsymbol{\mathrm{G}}( \boldsymbol{\mathrm{r}}_2 , \boldsymbol{\mathrm{r}}_1 )  ) =  0, ~ \mathrm{Re}( \boldsymbol{\mathrm{G}}( \boldsymbol{\mathrm{r}}_2 , \boldsymbol{\mathrm{r}}_1 )  ) = 0.9 $. In all cases the Green function quantity is normalized by $ \mathrm{Im}( \boldsymbol{\mathrm{G}}( \boldsymbol{\mathrm{r}}_1 , \boldsymbol{\mathrm{r}}_1 )  ).$}\label{diss_coherent}
	\end{center}
\end{figure}

In order to demonstrate this difference between reciprocal and unidirectional systems, we consider two cases of pure dissipative and pure coherent qubit communication for a model system where we simply assign the Green function values. Fig. \ref{diss_coherent} shows that the pure dissipative regime is dominant for the reciprocal case while the dissipative or coherent nature of the qubit communication is unimportant for the unidirectional case.  

%%%%%%%%%%%%%%%%%%%%%%%%%%%%%%%%%%%%%%%%%%%%%%
%%%%%%%%%%%%%%%%%%%%%%%%%%%%%%%%%%%%%%%%%%%%%%
\section{Numerical results}\label{NR}
A unidirectional SPP can be provided by the interface between a PTI and a topologically-trivial material. When operated in a common bandgap of the two materials (or if the trivial medium is opaque), the SPP is unidirectional, topologically protected from back-scattering, and diffraction-immune, providing an ideal implementation of a strongly nonreciprocal system for qubit interactions. Although here we implement a PTI as a PQHE using a continuum plasma \cite{Mario2}-\cite{Hassani3}, many other implementations of PTIs are possible, of both PQHE and PQSHE types, and qualitatively would behave in a similar manner. 

\subsection{Continuum photonic topological insulator realization of a nonreciprocal surface plasmon polariton environment}
We assume a magnetized plasma having the permittivity tensor
\begin{equation}\label{Eq:plasma_epsilon}
\underline{\varepsilon}=\left[ 
\begin{array}{ccc}
\varepsilon_{11} & i \varepsilon_{12} & 0 \\ 
-i \varepsilon_{12} & \varepsilon_{11} & 0 \\ 
0 & 0 & \varepsilon_{33}%
\end{array}%
\right]  
\end{equation}% 
where
\begin{align}
& \varepsilon_{11} = 1 + i \frac{\omega_p^2}{\omega}\left( \frac{  \nu - i \omega }{(\nu - i \omega)^2 + \omega_c^2} \right) \notag \\&
\varepsilon_{12} = \frac{ \omega_p^2 \omega_c }{ \omega \left( (\nu - i \omega)^2 + \omega_c^2  \right) },\,
\varepsilon_{33} = 1 + i \frac{\omega_p^2}{ \omega (\nu - i \omega) },
\end{align}
and where $\omega _{c}=\left( q_{e}/m_{e}\right)\mathrm{B}_{z}$ is the cyclotron frequency ($\mathrm{B}_{z}$ is the applied bias field), $\omega _{p}^{2}=N_{e}q_{e}^{2}/\varepsilon _{0}m_{e}$ is the squared plasma frequency ($N_{e}$ is the free electron density and $q_{e}$ and $m_{e}$ are the electron charge and mass, respectively), and $ \nu $ is the collision frequency. Initially, we set $ \nu = 0 $ to focus on the effect of unidirectionality, but later the effect of loss is considered. The magnetized plasma is able to support a bulk TE mode with dispersion $ k^2_{\mathrm{TE}} = \varepsilon_{33} (\omega/c)^2 $ and a bulk TM mode with dispersion $ k^2_{\mathrm{TM}} = \varepsilon_{\mathrm{eff}} (\omega/c)^2 $, where $ \varepsilon_{\mathrm{eff}} = ( \varepsilon^2_{11} - \varepsilon^2_{12} )/\varepsilon_{11} $ \cite{Arthur}. Both bulk modes are reciprocal. The Chern number of the bulk TE mode is trivial, and so TE modes are not considered further in this work. The Chern numbers of the bulk TM modes are nonzero, and at the interface of the magnetized plasma and a topologically-trivial (simple) medium the gap Chern number is $\mathrm{C_{gap}=1}$ \cite{Mario2,Hassani2}, indicating the presence of one nonreciprocal, backscattering-immune TM-SPP that crosses the bandgap (bandstructure is shown later, in Fig. \ref{DB}).  

%%%%%%%%%%%%%%%%%%%%%%%%%%%%%%%%%%%%%%%%%%%%%%
\subsection{Entanglement evaluation in different environments}
\begin{figure}[h]
	\begin{center}
		\noindent
		\includegraphics[width=3.5in]{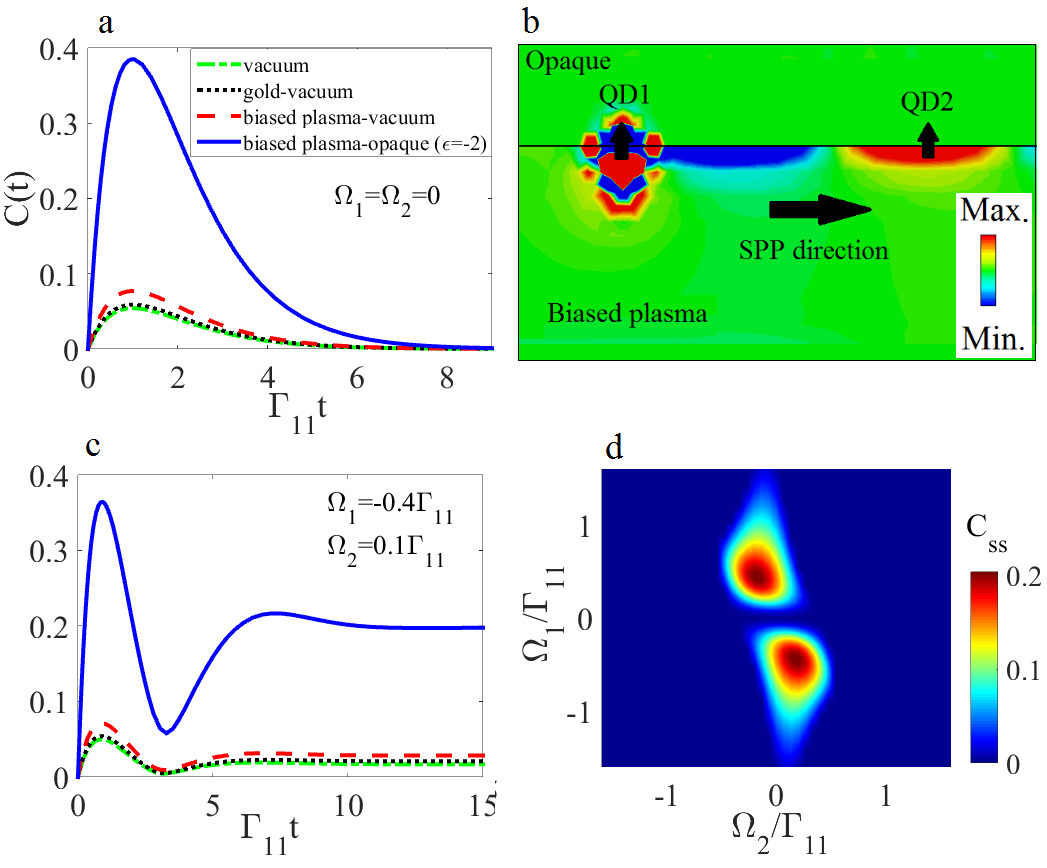}
		\caption{a. Transient concurrence for two interacting qubits in different environments; 1) vacuum, 2) at the interface of a gold half-space ($ \varepsilon = -91.6-3i  $) and vacuum, 3) at the interface of a magnetized plasma ($ \omega_p / \omega = 0.95,~\omega_c / \omega = 0.21  $) and vacuum, and 4) at the interface of the magnetized plasma and an opaque medium (non-biased plasma with $ \omega_p/\omega = \sqrt{3} $, such that $\varepsilon=-2$). b. One way SPP at the interface of the biased plasma and the opaque medium at $ \omega/2\pi=200 $ THz. c. Driven concurrence of two qubits in the same environments as in panel a. d. Steady states concurrence versus pumping intensities for the case of the biased plasma and opaque medium interface. The qubit separation is $ 2.4 $ $ \mu $m ($1.6 \lambda_0$).}\label{Fig2_a}
	\end{center}
\end{figure}

We first consider the behavior of the concurrence for qubits in several different environments, and establish that the best entanglement occurs for a PTI/opaque medium interface. Fig. \ref{Fig2_a}a, shows a comparison of concurrence between four different cases of two qubits interacting through: 1) vacuum, 2) at the interface of a gold half-space and vacuum, 3) at the interface of a magnetized plasma and vacuum, and 4) at the interface of a magnetized plasma and an opaque medium. Here and in the following, the Green function is calculated numerically \cite{COMSOL}.
The system of qubits were initially prepared in state $\left|4\right> = \left|e_1\right> \otimes \left|g_2\right>$, such that the left qubit is initially in the excited state while the right qubit is in the ground state. It can be seen that the interface between the magnetized plasma and the opaque medium has higher concurrence then the other cases, due to the existence of a strong SPP and the fact that there can be no radiation into either bulk half-space. Thus, in the following, we focus on the magnetized plasma/opaque medium geometry. 

Figure \ref{Fig2_a}b shows the excited unidirectional SPP at the interface of the magnetized plasma and the opaque medium, demonstrating the unidirectional nature of the SPP, and Fig. \ref{Fig2_a}c shows the case of pumped concurrence, where the qubit depopulation is compensated by applying an external laser source in resonance with the atomic transition frequency. The pump intensity must be chosen carefully, as illustrated in Fig. \ref{Fig2_a}d, which shows the steady state concurrence for a wide range of laser intensities (a laser pump can be applied to the qubits via, e.g., a fiber penetrating into the material). It can be seen that the laser intensity can not be too large, otherwise the qubits will interact mostly with the laser. Ideally, the pump should be strong enough to keep the system interacting, but weak enough for the qubit interaction to dominate the dynamics. It is clear from Fig. \ref{Fig2_a}d that unequal pumping leads to larger steady state concurrence.   

\subsection{Topological aspect of entanglement}
\begin{figure}[h]
	\begin{center}
		\noindent
		\includegraphics[width=3.5in]{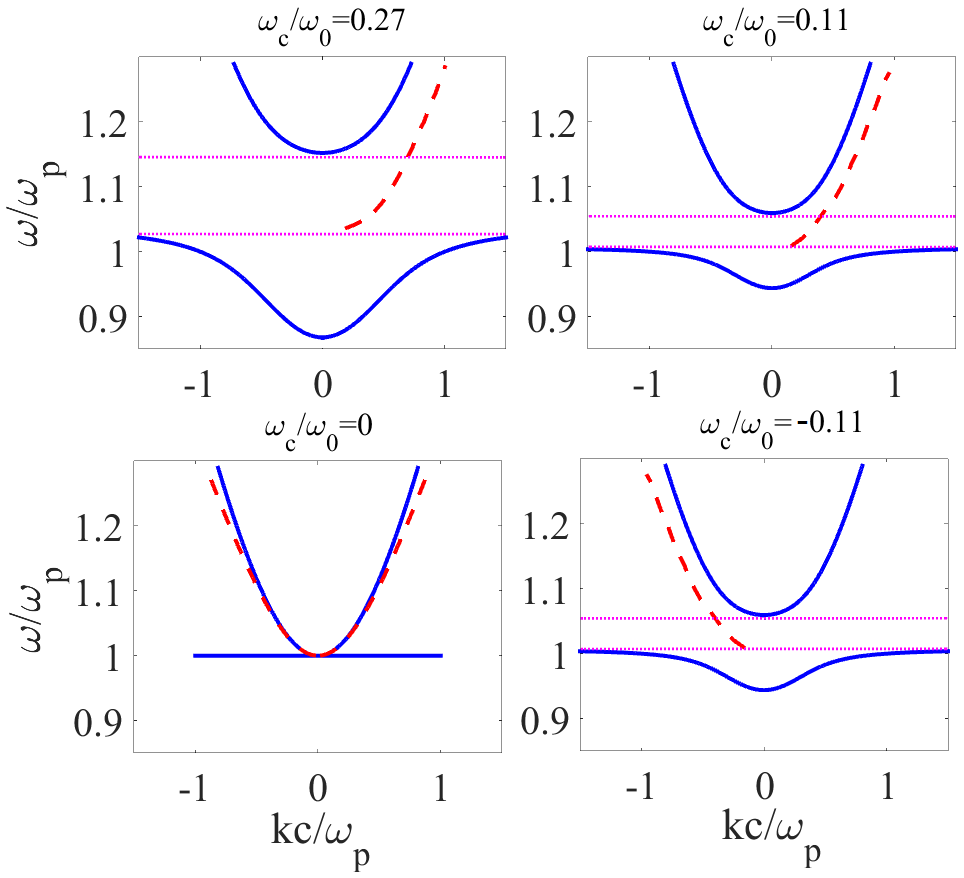}
		\caption{Reciprocal bulk bands (solid blue) for the biased plasma ($ \omega_p/\omega = 0.95 $), and the unidirectional gap-crossing SPP (dashed red) dispersion for a biased-plasma/opaque medium ($\varepsilon=-2$) interface, for different values of bias at $ \omega/ 2 \pi = 200 $ THz. }\label{DB}
	\end{center}
\end{figure}

In this section we briefly show the topological aspect of entanglement in a PTI system. Figure \ref{DB} shows the reciprocal bulk bands (solid blue) for the biased plasma, and the unidirectional gap-crossing SPP (dashed red) dispersion for a biased-plasma/opaque medium interface, for different values of bias. For $ \omega_c>0$ the gap Chern number is -1 \cite{Mario2,Hassani2}, and there is a positive-traveling SPP ($v_g=d\omega/dk>0$), topologically-protected against backscattering. At $\omega_c=0$ the gap closes, the material becomes topologically-trival (gap Chern number is 0), and there exists a reciprocal SPP. For $ \omega_c<0$ the gaps reopens, the gap Chern number is 1, and there is a negative-traveling SPP ($v_g<0$), topologically-protected against backscattering. 

\begin{figure}[h]
	\begin{center}
		\noindent
		\includegraphics[width=2.8in]{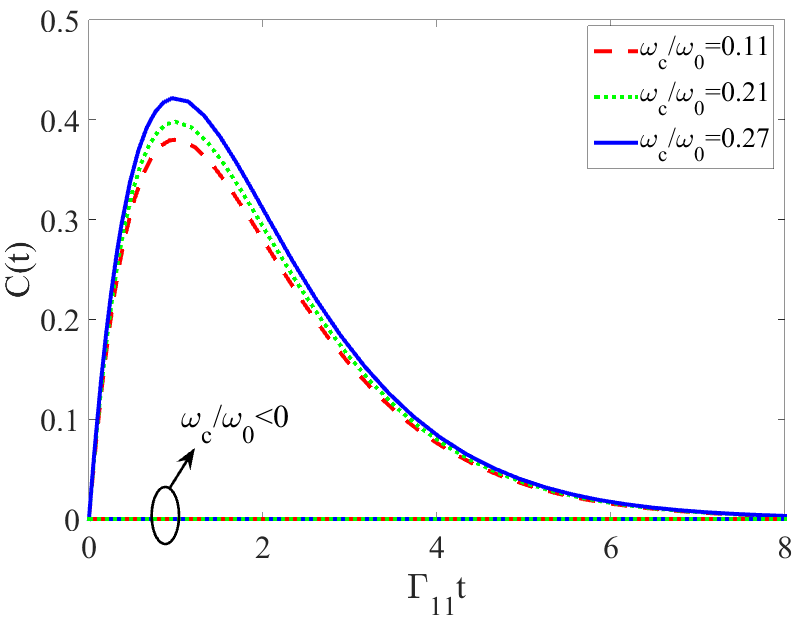}
		\caption{Concurrence mediated by a unidirectional SPP at the interface of biased plasma ($ \omega_p/\omega = 0.95 $) and an opaque medium ($\varepsilon=-2$) when the left dot has an initial excitation (state $ \left| 4 \right> = \left| e_1,g_2 \right>$). For $\omega_c<0$, the same three absolute values are considered as for positive bias, i.e., $\omega_c/\omega_0=-|0.27|, -|0.21|$, and $-|0.11|$. The qubit separation is $ 2.4 $ $ \mu $m ($1.6 \lambda_0$).}\label{Ctf}
	\end{center}
\end{figure}

Figure \ref{Ctf} shows the concurrence when the left dot has an initial excitation (state $ \left| 4 \right> = \left| e_1,g_2 \right>$). The concurrence is rather insensitive to the bias as long as the topology does not change, however, when the gap closes and reopens the concurrence vanishes.

\subsection{Preserving entanglement in the presence of large defects}

Perhaps the most important aspect of using PTIs for entanglement is the possibility of robust SPPs, topologically-immune to backscattering (and immune to diffraction if operated in the bulk bandgap) in the presence of any arbitrary large obstacle or defect. To examine this, we compare two cases: 1) the interface between an opaque medium and a biased plasma, and 2) the interface between the same opaque medium and an unbiased plasma.  

\begin{figure}[h]
	\begin{center}
		\noindent
		\includegraphics[width=3.5in]{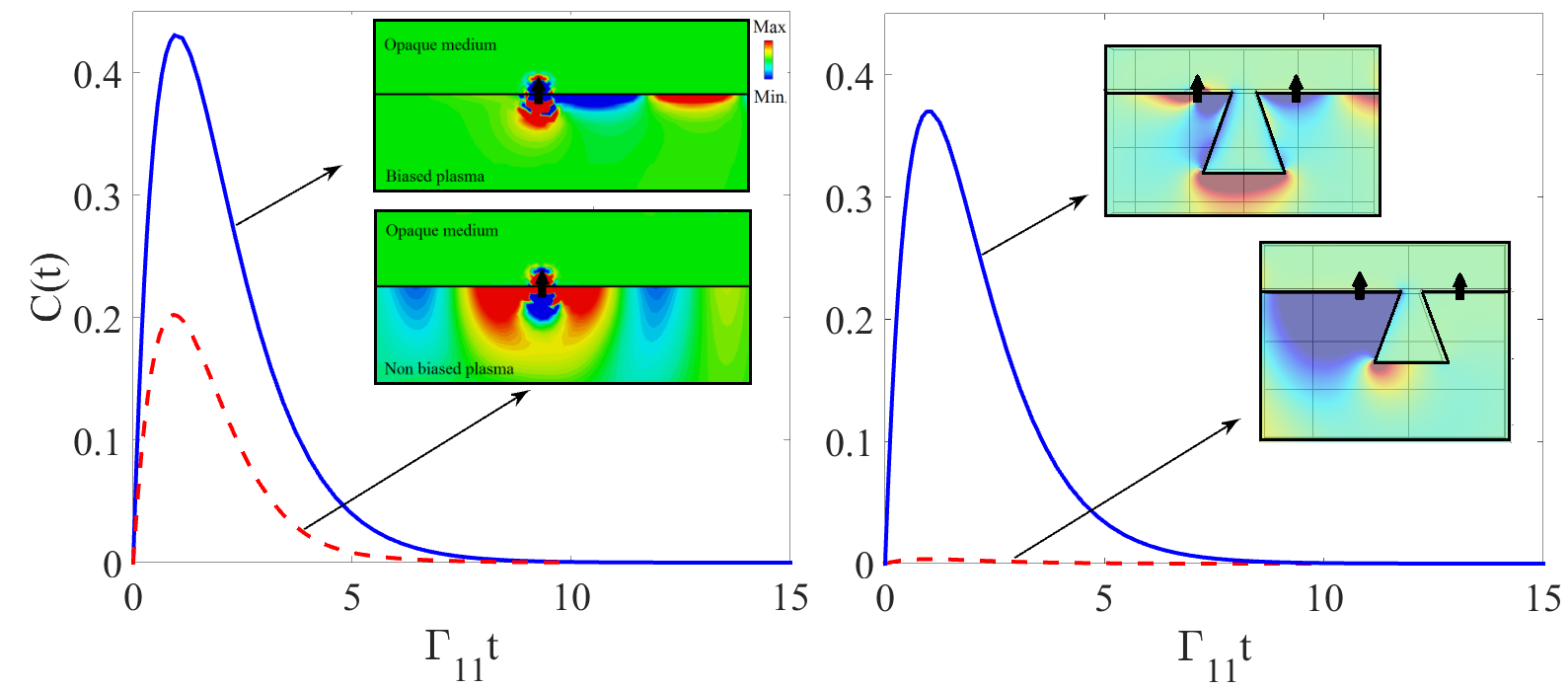}
		\caption{Left panel: Transient concurrence of two qubits interacting through a flat interface made of an opaque medium ($\varepsilon=-2$) and both an unbiased plasma ($  \omega_p / \omega = 0.95 $, $  \omega_c / \omega = 0$) and a biased plasma ($ \omega_p / \omega = 0.95 $, $  \omega_c / \omega = 0.21$). Insert shows shows the electric field $ \mathrm{E}_y $ excited by a vertical electric dipole.  Right panel: Same thing for the case of a defected interface, where the defect contour length is of the order of a free-space wavelength. The system of qubits is initially prepared in the state $ \left| 4 \right> = \left| e_1,g_2 \right>$. The qubit separation is $ 1.7 $ $ \mu $m ($1.13 \lambda_0$).}\label{defected}
	\end{center}
\end{figure}

In the nonreciprocal case, this unidirectional and scattering-immune SPP provides the ability to preserve the entangled state of two qubits in plasmonic systems even in the presence of very non-ideal interfaces. Figure \ref{defected} shows the transient concurrence for the cases of biased/unbiased plasmas with flat and defected interfaces. Although for the flat interface the biased plasma provides better concurrence then the reciprocal (unbiased) case, this could be perhaps altered by adjustment of the two material half-space properties. However, the point is that in the presence of a defect, as shown in the right panel, the reciprocal SPP suffers from a strong reflection at the defect, as expected, whereas the nonreciprocal SPP (biased plasma) detours around the defect, leading to the same concurrence as without the defect. 

\subsection{Finite-width waveguide}
\begin{figure}[h]
	\begin{center}
		\noindent
		\includegraphics[width=3.5in]{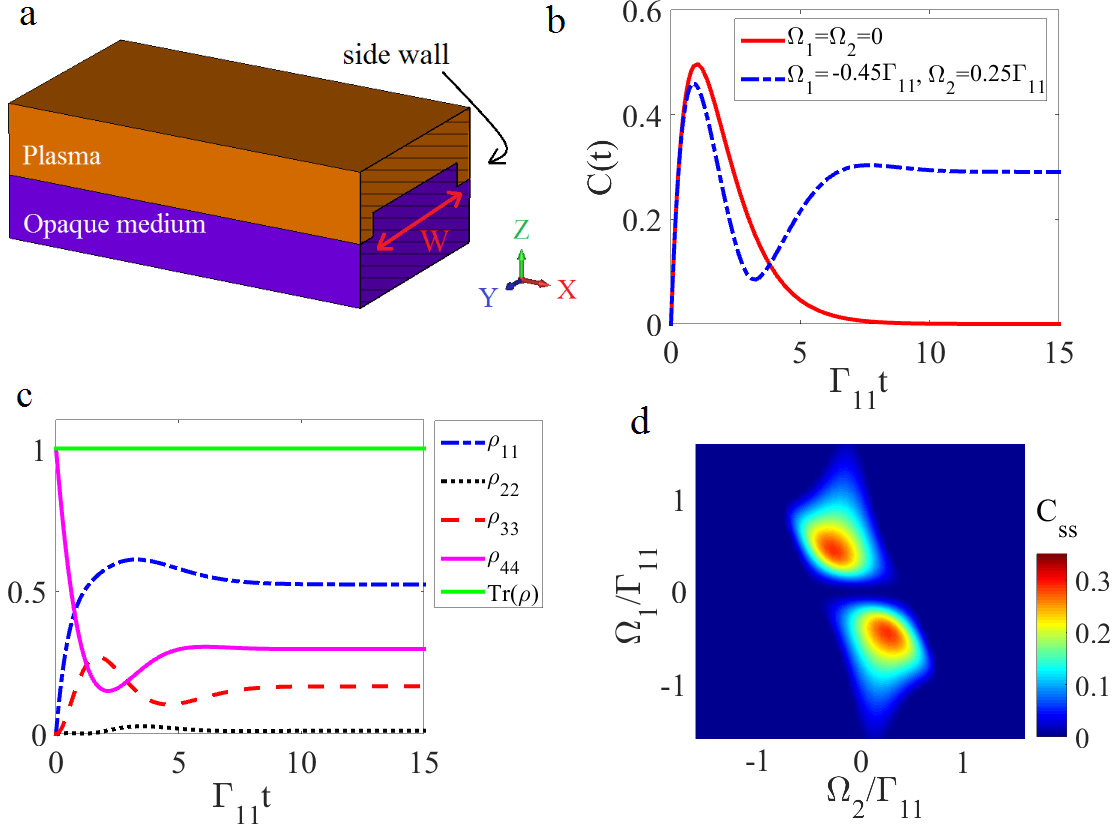}
		\caption{a. Finite-width waveguide formed by an opaque medium and biased plasma. b. Transient and driven concurrence of two qubits interacting through the finite-width waveguide. For the biased plasma, $ \omega_p / \omega = 0.95 $ and  $  \omega_c / \omega = 0.21$, and for the opaque medium, $\varepsilon=-2$. c. Dynamics of the qubits under external pumping. d. Steady state concurrence for different pump values. Waveguide width is $ 1.8 $ $\mu$m ($1.2 \lambda_0$) and qubit separation is 2.4 $\mu$m ($1.6 \lambda_0$)}.\label{Fig4}
	\end{center}
\end{figure}

The previous results were for an infinitely-wide interface. In this section we examine the effect of lateral confinement of the SPP \cite{Hassani2}. Figure \ref{Fig4}a shows the finite-width waveguide geometry. In order to efficiently confine the SPP along the propagation axis, the plasma is extended past the interface to form partially-extended sidewalls. Only partial side walls are needed to prevent radiation in space, since the SPP is confined to the vicinity of the interface. 

Lateral confinement of the unidirectional SPP improves both the transient and steady state (pumped) concurrence. Fig. \ref{Fig4}b shows the transient and steady state concurrence of two qubits initially prepared in state $ \left| 4 \right >$. In comparison to Fig. \ref{Fig2_a}a, it can be seen that lateral confinement increases both the maximum transient concurrence and the steady state concurrence. Figure \ref{Fig4}c shows the dynamics of the qubits under external pumping, where $ \rho_{11}, ~ \rho_{22}, ~ \rho_{33},~\mathrm{and} ~ \rho_{44}  $ are the probabilities of finding both qubits to be in ground state, both qubits in the excited state, the first qubit in the ground state and the second qubit in the excited state, and vice versa, respectively. Figure \ref{Fig4}d shows the steady state concurrence for a wide range of pump values. The behavior is similar to the case of the infinite interface, Fig. \ref{Fig2_a}d, except that the range of pump values that result in large steady state concurrence is extended, and the maximum achievable steady state concurrence is larger in the case of the finite-width waveguide.

\begin{figure}[h]
	\begin{center}
		\noindent
		\includegraphics[width=3in]{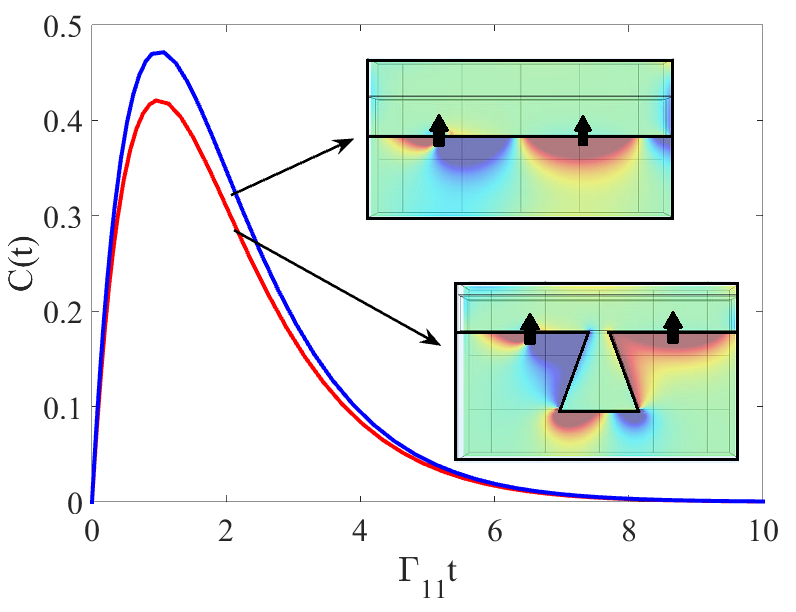}
		\caption{Transient concurrence of two qubits interacting in a finite-width waveguide (see Fig. \ref{Fig4}a) consisting of an opaque medium ($\varepsilon=-2$) and a biased plasma ($ \omega_p / \omega = 0.95 $, $  \omega_c / \omega = 0.21$). The defect contour length is of the order of a free-space wavelength, and spans the width of the waveguide, $W=1.8$ $\mu$m ($1.2 \lambda_0$). Qubit spacing for the flat interface is 2.4 um ($1.6 \lambda_0$), and for the interface with defect, the line-of-sight spacing is 2.4 um. The system of qubits is initially prepared in the state $ \left| 4 \right> = \left| e_1,g_2 \right>$. }\label{FWD}
	\end{center}
\end{figure}

In Fig. \ref{FWD}, qubit concurrence is shown for a finite-width waveguide having a defect which spans the entire waveguide width. It can be seen that the concurrence is minimally affected by the defect. Although not shown, as with Fig. \ref{defected}, in the reciprocal (unbiased) case the defect eliminates the concurrence.

\subsection{Effect of different initial state preparations}
\begin{figure}[h]
	\begin{center}
		\noindent
		\includegraphics[width=3.45in]{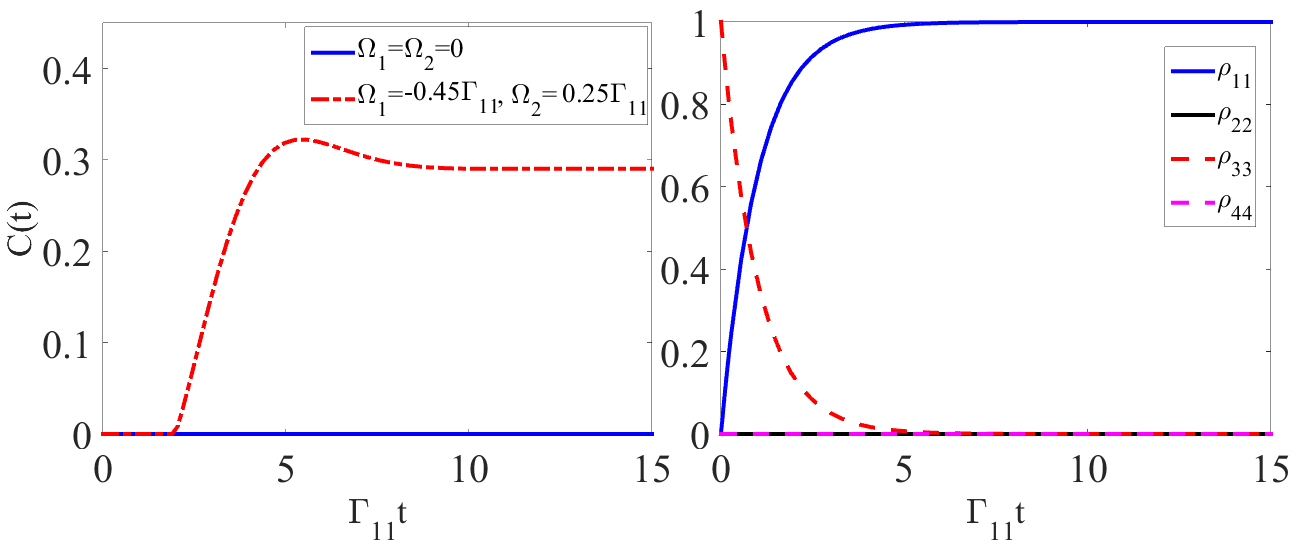}
		\caption{Left panel: Transient and driven concurrence for a system of qubits interacting through a right going unidirectional SPP while the initial excitation is in the right qubit. Right panel: Dynamics of the qubit system for the transient case. For the biased plasma $ \omega_p / \omega = 0.95 $ and  $  \omega_c / \omega = 0.21$, and for the opaque medium $\varepsilon=-2$. The waveguide geometry is shown in Fig. \ref{Fig4}a, and qubit separation is 2.4 $\mu$m ($1.6 \lambda_0$)}\label{Fig5}
	\end{center}
\end{figure}

An interesting behavior of the concurrence arising from having a unidirectional SPP is that, e.g., if the medium supports only a right going SPP, then the initially excited qubit should be the left qubit, otherwise the qubits remain unentangled, as shown in Fig. \ref{Fig5}a for the unpumped case $\Omega_1=\Omega_2=0$. Figure \ref{Fig5}b shows the dynamics of the qubits for this unpumped case. It can be seen that $ \rho_{33} $, which is the probability of finding the right qubit in the excited state and the left qubit in the ground state, starts from 1 and then drops rapidly. However, $ \rho_{44} $, which is the probability of finding the excitation being in the left qubit with the right qubit in the ground state, is always zero, meaning that the excitation lost from the right qubit never gets captured by the left qubit. This behavior is particular to a unidirectional environment, and allows for keeping two qubits disentangled at any qubit separation, even if one of them carries an excitation. 

However, by applying an external pump we can achieve non-zero concurrence, as also depicted in Fig. \ref{Fig5}a
The pump is turned on at $\mathrm{t}=0$, and instead of immediately becoming non-zero, the concurrence remains zero for a period of time, then starts raising as a sudden birth in concurrence and reaches a non-zero steady state value. This delayed sudden-birth is quite different from the pumped reciprocal case. 

\begin{figure}[h]
	\begin{center}
		\noindent
		\includegraphics[width=3.35in]{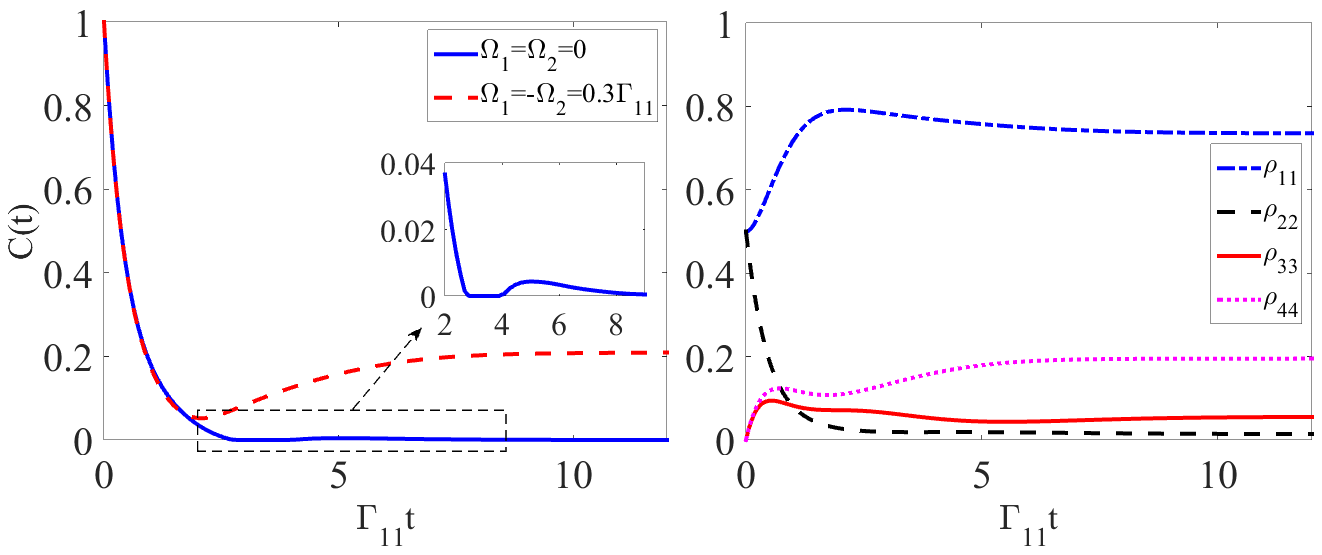}
		\caption{Left panel: Transient and driven concurrence for a system of qubits initially prepared in the Bell state. Right panel: Dynamics of the qubits system under external pumping. For the biased plasma, $ \omega_p / \omega = 0.95 $ and  $  \omega_c / \omega = 0.21$, and for the opaque medium $\varepsilon=-2$. The waveguide geometry is shown in Fig. \ref{Fig4}a with $ W=1.8 $ $\mu$m ($1.2 \lambda_0$), and qubit separation is 2.4 $\mu$m ($1.6 \lambda_0$).} \label{Bell}
	\end{center}
\end{figure}

It is also possible to consider different initial states which can give other possible unidirectional SPP assisted dynamical evolutions. Figure \ref{Bell} shows the case of the initial state being the maximally entangled Bell state $ \left| \Psi_{\mathrm{Bell}} \right > =  (\left| 1 \right> + \left| 2 \right>)/\sqrt{2} $. 
We consider that the qubits are interacting through the finite-width waveguide depicted in Fig. \ref{Fig4}a. Figure \ref{Bell}a shows the time evolution of the concurrence for both pumped and non-pumped cases. In contrast to the previous cases, the concurrence starts from one due to the maximum degree of entanglement of the initial Bell state. For the non-pumped case the concurrence diminishes in time as the system becomes disentangled, resulting in a sudden death of entanglement. It remains zero for a period of time, then the entanglement experience a rebirth before decaying exponentially at long times. For the externally pumped case, the concurrence exponentially decays but the qubits do not become completely disentangled. Fig. \ref{Bell}b shows the dynamics of the qubits for the pumped case. The population probabilities $ \rho_{11} $ and $ \rho_{22} $ start from 0.5 due to the Bell state preparation. An interesting behavior in the qubit dynamics is the unequal steady state values $ \rho_{33} $ and $ \rho_{44} $ values under pumping with equal intensities $ |\Omega_1| = |\Omega_2| $ (in the reciprocal case, $ \rho_{33} = \rho_{44} $). 
%%%%%%%%%%%%%%%%%%%%%%%%%%%%%%%%%%%%%%%%%%%%%%
%%%%%%%%%%%%%%%%%%%%%%%%%%%%%%%%%%%%%%%%%%%%%%
\subsection{Lossy biased plasma}
\begin{figure}[h]
	\begin{center}
		\noindent
		\includegraphics[width=3.5in]{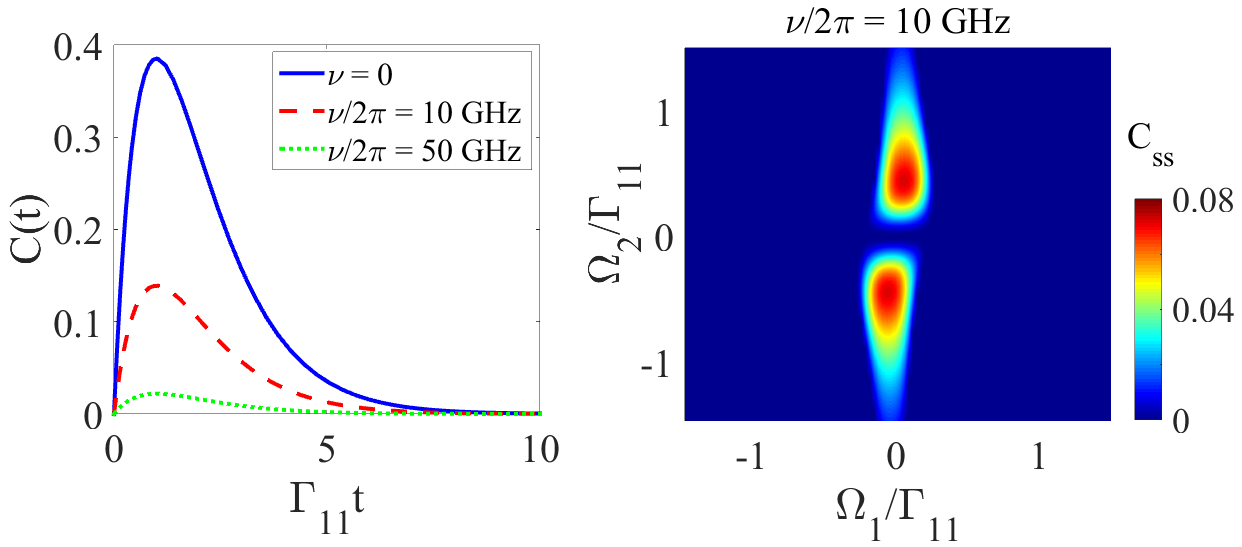}
		\caption{Left panel: Transient concurrence of two qubits interacting through an infinite interface between a biased plasma ($ \omega_p / \omega = 0.95 $ and  $  \omega_c / \omega = 0.21$) and an opaque medium ($\varepsilon=-2$) for different values of the collision frequency. Right panel: Steady state concurrence for different pump values in the lossy case. Qubit separation is 2.4 $\mu$m ($1.6 \lambda_0$)}\label{lossy}
	\end{center}
\end{figure}

In a lossy medium the SPP loses power as it propagates along the interface, resulting in weaker qubit entanglement. In order to study the effect of loss, we suppose the qubits are interacting through an infinitely-wide interface as considered in Fig. \ref{Fig2_a}, but for three different collision frequencies; $ \nu = 0 $, $ \nu / 2 \pi = 270  $ MHz and $ \nu / 2\pi = 500  $ MHz. Qubits are initially prepared in the state $\left|4\right> = \left|e_1\right> \otimes \left|g_2\right>$. Figure \ref{lossy}, left panel, shows the transient concurrence. Increasing the collision frequency reduces the concurrence, and for collision frequencies greater than 500 MHz loss dominates the system and an entangled state is not achievable for this relatively wide qubit separation of $1.6 \lambda_0$. 

%From (\ref{Eq:Ct}) it can be seen that $ \Gamma_{11} $ determines how rapidly the concurrence decays. The key parameter for large entanglement is having large $ \boldsymbol{\mathrm{G}}(\boldsymbol{\mathrm{r}}_i, \boldsymbol{\mathrm{r}}_j) / \mathrm{Im} \boldsymbol{\mathrm{G}}(\boldsymbol{\mathrm{r}}_i, \boldsymbol{\mathrm{r}}_i) $. For simplicity, we assume qubits at the interface between the two materials, which also results in strong coupling of the qubits to the SPP. However, as we increase the material loss, $ \Gamma_{ii} $ increases due to strong quenching, which leads to a smaller ratio of $ \boldsymbol{\mathrm{G}}(\boldsymbol{\mathrm{r}}_i, \boldsymbol{\mathrm{r}}_j) / \mathrm{Im} \boldsymbol{\mathrm{G}}(\boldsymbol{\mathrm{r}}_i, \boldsymbol{\mathrm{r}}_i) $ and weaker concurrence. For a given value of loss, the position of the qubits above the interface could be tuned to achieve the best possible concurrence, although when the qubits are positioned further from the interface.  

The right panel of Fig. \ref{lossy} shows the steady state concurrence of the pumped system, versus pumping intensity. In comparison to the lossless case (Fig. \ref{Fig2_a}d), the range of pump intensities that give non-zero steady state concurrence has decreased, and the maximum achievable concurrence value is diminished compared to the lossless case.
%%%%%%%%%%%%%%%%%%%%%%%%%%%%%%%%%%%%%%%%%%%%%%
%%%%%%%%%%%%%%%%%%%%%%%%%%%%%%%%%%%%%%%%%%%%%%
%%%%%%%%%%%%%%%%%%%%%%%%%%%%%%%%%%%%%%%%%%%%%%

\section{Conclusions}

We have derived a master equation for qubit dynamics in a general three-dimensional, nonreciprocal, inhomogeneous and lossy environment. Spontaneous and pumped entanglement were investigated for two qubits in the vicinity of a photonic topological insulator interface, which supports a nonreciprocal (unidirectional), scattering-immune surface plasmon polariton in the bandgap of the bulk material. We have illustrated the topological nature of the entanglement, and it was shown that large defects in the interface do not impact entanglement for the PTI case, whereas a defect has considerable effect for the reciprocal case. Several initial qubit states were considered, as well as the influence of pump intensity and material loss. Particularities arising from the unidirectional nature of the qubit communication were highlighted.

\section*{Acknowledgments}

The authors would like to thank Stefan Buhmann for helpful discussions. 

\section*{APPENDIX}

\section*{Appendix I: Master equation approximations}\label{AI}
Here we briefly discuss the approximations used in the derivation of the master equation (\ref{Eq:ME})-(\ref{Eq:Lindblad_bi}). 

The first approximation made in the derivation is the rotating wave approximation (RWA) where in the interaction picture we drop the rapidly varying counter-rotating terms in $ \mathrm{H}_I $. This approximation is valid for $ \Gamma_{ii} \ll \omega $. The qubit transition frequency is $ \omega / 2\pi = 200 $ THz, and we assume a dipole moment $ \mathrm{d} = 60 $ D. For an interface made of lossless biased plasma, $ \Gamma_{ii} / 2\pi \sim 450$ MHz, and for the lossy biased plasma with $ \nu / 2\pi = 500 $ MHz, $ \Gamma_{ii} / 2\pi \sim 2$ GHz. For the non biased plasma-opaque medium interface (interface supporting reciprocal SPP) $ \Gamma_{ii} / 2\pi \sim 75$ MHz. In all cases the condition for the validity of the RWA is strongly met.

We also applied the Born-Markov approximation, which comes from the assumption that the reservoir relaxation time, $ \tau_R $, is much faster than the relaxation time of the qubit system $ \tau_S=  1/\Gamma_{ii} $. This allows for the expansion of the exact equation of motion for the density matrix up to second order, and makes the quantum master equation local in time. For a nonreciprocal medium the fluctuation-dissipation theorem \cite{Buh} is 
\begin{align}
	\left< \mathrm{P}_{\alpha}( \omega, \boldsymbol{\mathrm{r}} ) \mathrm{P}^{\dagger}_{\alpha}( \omega', \boldsymbol{\mathrm{r}}' ) \right>  =  &\frac{\hbar}{4i} ( \boldsymbol{\varepsilon} ( \omega, \boldsymbol{\mathrm{r}} ) - \boldsymbol{\varepsilon}^{\dagger} ( \omega, \boldsymbol{\mathrm{r}} )  ) \mathrm{N}(\omega , \mathrm{T})  \notag \\&
	\times \delta (\omega - \omega') \delta( \boldsymbol{\mathrm{r}} - \boldsymbol{\mathrm{r}}'  ) \delta_{\alpha\beta},
\end{align}
where $ \mathrm{N}(\omega, \mathrm{T}) = 2/\left( \mathrm{exp}(\hbar \omega / k_B \mathrm{T}) - 1 \right) $ for negative frequencies and $ \mathrm{N}(\omega, \mathrm{T}) = 1+  2/\left( \mathrm{exp}(\hbar \omega / k_B \mathrm{T}) - 1 \right) $ for positive frequencies, where $k_B$ is Boltzmann's constant. Regarding $ \boldsymbol{\mathrm{E}}(\mathbf{r},\omega)  = (\omega^2 / c^2 \varepsilon_0) \int_V d \mathbf{r}' \boldsymbol{\mathrm{G}}(\mathbf{r}, \mathbf{r}', \omega) \cdot \boldsymbol{\mathrm{P}}(\mathbf{r}' , \omega)  $, it can be shown that 
\begin{align}
	\left <  \mathrm{E}_{\alpha} (\boldsymbol{\mathrm{r}},\omega) \mathrm{E}_{\alpha}^{\dagger} (\boldsymbol{\mathrm{r}},\omega) \right> & = \mathrm{k}^2_0 \frac{\hbar}{4i \varepsilon^2_0} \mathrm{N}(\omega, \mathrm{T}) \notag \\&
	\times (\mathrm{G}_{{\alpha}{\alpha}} (\boldsymbol{\mathrm{r}}, \boldsymbol{\mathrm{r}}, \omega ) - \mathrm{G}^{\dagger}_{{\alpha}{\alpha}} (\boldsymbol{\mathrm{r}}, \boldsymbol{\mathrm{r}}, \omega )),
\end{align}
which reduces in the reciprocal case to 
\begin{align}
	\left <  \mathrm{E}_{\alpha} (\boldsymbol{\mathrm{r}},\omega) \mathrm{E}_{\alpha}^{\dagger} (\boldsymbol{\mathrm{r}},\omega) \right> & = \frac{\hbar \mathrm{k}^2_0}{2\varepsilon^2_0}  \mathrm{N}(\omega, \mathrm{T})
	\mathrm{Im}(\mathrm{G}_{{\alpha}{\alpha}} (\boldsymbol{\mathrm{r}}, \boldsymbol{\mathrm{r}}, \omega )).
\end{align}
The bath relaxation time can be estimated by looking at the decay time of the correlation
\begin{equation}
	\left <  \mathrm{E}_{\alpha} (\boldsymbol{\mathrm{r}},t) \mathrm{E}_{\alpha}^{\dagger} (\boldsymbol{\mathrm{r}},0) \right>  = \frac{1}{2 \pi }  \int_{-\infty}^{+\infty} d \omega e^{-i \omega t}  \left <  \mathrm{E}_{\alpha} (\boldsymbol{\mathrm{r}},\omega) \mathrm{E}_{\alpha}^{\dagger} (\boldsymbol{\mathrm{r}},\omega) \right>.
\end{equation}
 The Green function consists of homogeneous (vacuum) and scattered terms, and $ \tau_R $ will be dominated by the slower scattered field contribution (for the vacuum term, $ \tau_R(T)=\hbar/\pi k_B T$ \cite{BP}, so that $\tau_R(300K)\sim 10$ fs). For the scattered part of the Green function for an interface made of non-biased plasma-opaque medium (interface supporting reciprocal SPP), using the Green function in \cite{DN}, $ \tau_R \sim 10^{-11} $ s for $\nu=500$ MHz and  $\nu=270$ MHz, whereas $ \tau_S = 1/\Gamma_{ii} \sim 10^{-8} $ s, so that we can ignore the reservoir relaxation time.

%Finally, for large qubit separations or coupling rates, $ \Gamma_{ii}d \gg v_g $, non-Markovian effects arise in waveguide systems \cite{Garcia2}. In our case, as a rough estimate, considering a biased plasma-perfect metal interface $ \Gamma_{ii} \sim 10^8  $ and qubit separation is on the order of $ d \sim 10^{-6} $ so $ \Gamma_{ii}d \sim 10^2 $, whereas it can be shown that $ v_g \sim 10^8 <c $ where $c$ is the vacuum speed of light. Thus the condition $ \Gamma_{ii}d \ll v_g $ is strongly satisfied in our case.

\section*{Appendix II: Comparison with previous 1D chiral theory}\label{AII}
Here we discuss the relation between the general ME we derived in terms of the exact electromagnetic Green function, resulting, for two qubits, in the Lindblad \eqref{OME}, and the 1D phenomenological ME for two level systems coupled to a 1D chiral reservoir presented in \cite{Garcia2, Zoller2} (see also \cite{Zoller3, Zoller1}). The 1D chiral theory is based on the notion of right and left, defining couplings $\gamma_\textrm{R,L}$, whereas the theory presented here is based on qubit interactions $\Gamma_{ij}$; note that $\Gamma_{ij}$ plays the role of a $\Gamma_\textrm{right}$ if $x_i>x_j$, but plays the role of $\Gamma_\textrm{left}$ if $x_i<x_j$. To facilitate the comparison with the 1D chiral theory we will assume two qubits with positions $x_1$ and $x_2$, with $x_2>x_1$. In \cite{Garcia2,Zoller2}  phenomenological quantities $\gamma_{iR}, \gamma_{1L}$ for $i=1,2$ are utalized, and setting $ \gamma_{1R} = \gamma_{2R}=\gamma_R $ and $ \gamma_{1L} = \gamma_{2L}=\gamma_L $, the 1D chiral Lindblad superoperator is 
\begin{align}\label{TME}
	& \mathcal{L} {\rho}_{s}(t) = \sum_{j=1,2} \gamma_j \left(2 \sigma_j \rho_s \sigma^{\dagger}_j - \rho_s \sigma^{\dagger}_j \sigma_j - \sigma^{\dagger}_j \sigma_j \rho_s  \right) \notag \\ &
	+ \gamma_{R} e^{ik_R(x_2-x_1)} \left( \sigma_2 \rho_s \sigma^{\dagger}_1 - \rho_s \sigma^{\dagger}_1 \sigma_2 \right) \notag \\ & + \gamma_{R} e^{-ik_R(x_2-x_1)} \left( \sigma_1 \rho_s \sigma^{\dagger}_2 - \sigma^{\dagger}_2 \sigma_1 \rho_s \right) \notag \\&
	+  \gamma_{L} e^{-ik_L(x_2-x_1)} \left( \sigma_1 \rho_s \sigma^{\dagger}_2 - \rho_s \sigma^{\dagger}_2 \sigma_1 \right) \notag \\ & + \gamma_{L} e^{ik_L(x_2-x_1)} \left( \sigma_2 \rho_s \sigma^{\dagger}_1 - \sigma^{\dagger}_1 \sigma_2 \rho_s \right),
	\end{align}
where $k_{L,R}=\omega_0/v_{gL,R}$, with $v_g$ being the group velocity of the guided photons.

If we assume now a plasmonic environment, the total emission of the source can be divided into several decay channels: $ \Gamma_{11} = \Gamma_\mathrm{r} + \Gamma_{\mathrm{nr}} + \Gamma_{\mathrm{SPP}} $, where $\Gamma_{\mathrm{r}}$ represents free-space radiation, $ \Gamma_{\mathrm{nr}} $ represents losses in the material (quenching),  and $\Gamma_{\mathrm{SPP}}$ represents excitation of SPPs. Material absorption and radiation do not contribute to strong qubit-qubit interactions, and therefore we are interested in systems with strong decay through the plasmon channel, $ \Gamma_{\mathrm{SPP}} $, where the fraction of all emissions that are coupled to plasmons is expressed by $ \beta_{ij} = \Gamma_{ij,\mathrm{SPP}}/\Gamma_{11} $, with $i\neq j$. 

Assuming a plasmonic environment with a preferred propagation axis, here taken as $x$, in order to connect our formulation with previous 1D chiral formulations \cite{Garcia2, Zoller2} we introduce a particular 1D plasmonic version of (\ref{coupling_terms}), 
\begin{align}
& g_{ij}\simeq g_{ij,\mathrm{SPP}}={\beta _{ij}\Gamma _{11}}e^{-%
\mathrm{k}_{\mathrm{ij}}^{\prime \prime }|\mathrm{x}_{i}-\mathrm{x}_{j}|}%
\mathrm{sin}\left[ \mathrm{k}_{\mathrm{ij}}^{\prime }(\mathrm{x}_{i}-\mathrm{%
x}_{j})\right]   \nonumber  \label{Eq:pl1} \\
& \Gamma _{ij}\simeq \Gamma _{ij,\mathrm{SPP}}\\ &=\left\{ 
\begin{array}{c}
\left( \beta _{12}+\beta _{21}\right) \Gamma _{11},\ \ \ i=j\ \ \ \ \ \ \ \
\ \ \ \ \ \ \ \ \ \ \ \ \ \ \ \ \ \ \ \ \ \ \ \  \\ 
2\beta _{ij}\Gamma _{11}e^{-\mathrm{k}_{\mathrm{ij}}^{\prime \prime }|\mathrm{x}_{i}-\mathrm{x}_{j}|}\mathrm{cos}\left[ \mathrm{k}_{\mathrm{ij}%
}^{\prime }(\mathrm{x}_{i}-\mathrm{x}_{j})\right] ,\ \ \ i\neq j, \notag
\end{array}%
\right.
\end{align}%
where ${\mathrm{k}}_{\mathrm{ij}}={\mathrm{k}}_{\text{\textrm{spp},}\mathrm{%
ij}}=\mathrm{k}_{\mathrm{spp,ij}}^{\prime }+i\mathrm{k}_{\mathrm{spp,ij}%
}^{\prime \prime }$ are the SPP wavenumbers. In the systems considered here the bulk modes are reciprocal, whereas the interface SPP is strongly nonreciprocal (unidirectional). Thus, to compare with the 1D chiral ME it is sensible to consider the SPP (nonreciprocal) contribution. 

As defined in \eqref{Eq:pl1}, $\Gamma _{ij,\mathrm{SPP}}$ is discontinuous at $x_i=x_j$ in the nonreciprocal case, i.e., $\Gamma_{ij,\mathrm{SPP}}=2\beta_{ij}\Gamma_{11}$ as $|x_i \rightarrow x_j|$, whereas at $x_i=x_j$, $\Gamma_{ij,\mathrm{SPP}}=(\beta_{12}+\beta_{21}) \Gamma_{11}$. As we show below, the SPP contribution in the considered PTI system is indeed discontinuous at $x_i=x_j$. However, the exact $\Gamma_{ij}$, which contains both the SPP and radiation continuum, is continuous at the source point even in the nonreciprocal case. As another example of this, a 3D analytical Green function for a nonreciprocal bulk medium is provided in \cite{AL} (see their Eq. (117)), where $\Gamma_{ij}$ is also seen to be continuous.

Equating (\ref{OME}) in the 1D case (i.e., using (\ref{Eq:pl1})) and (\ref{TME}) term by term, the two Lindblad superoperators will be equal if 
\begin{eqnarray}
\gamma _{j} &=&\frac{\Gamma _{jj}}{2} \label{1a},\\
\gamma _{R}e^{\pm ik(x_2-x_1) } &=&\frac{\Gamma _{21}}{2}\pm ig_{21}, \label{1b}\\
\gamma _{L}e^{\pm ik(x_1-x_2)}&=&\frac{\Gamma _{12}}{2}\pm ig_{12}. \label{1c}
\end{eqnarray}
If we now make the assignments
\begin{eqnarray}
\beta _{21}\Gamma _{11} &\rightarrow &\gamma _{R},\ \ \ \beta _{12}\Gamma
_{11}\rightarrow \gamma _{L}, \label{2a} \\
{\mathrm{k}}_{\text{\textrm{spp},}\mathrm{12}} &\rightarrow &\frac{\omega
_{0}}{v_{gL}},\ \ \ {\mathrm{k}}_{\text{\textrm{spp},}\mathrm{21}%
}\rightarrow \frac{\omega _{0}}{v_{gR}}, \label{2b}
\end{eqnarray}
then \eqref{1a}-\eqref{1c} are satisfied and (\ref{OME}) becomes strictly equal to (\ref{TME}). It is worth stressing that physically the two formulations still differ, since \eqref{2b} is not exact (phase velocity and group velocity are different quantities). Nonetheless it is interesting to try to connect the phenomenological parameters in the model \eqref{TME} to the corresponding ones in \eqref{OME}, which are obtained in terms of the Green function, and hence can be computed for arbitrary environments. 

Using the rates defined in (\ref{Eq:pl1}), (\ref{Eq:Ct}) reduces to 
\begin{align}\label{Eq:Ct_spp}
\mathcal{C}^{\mathrm{1D}}(t) & =  2\beta_{21} \Gamma_{11} e^{-k''_{\mathrm{spp}}|x_2-x_1|}  t e^{-\Gamma_{11}t},
\end{align}
which is distance-independent in the lossless case, as noted in \cite{Garcia2} (using \eqref{2a}, \eqref{Eq:Ct_spp} is the same as Eq. (6) in \cite{Garcia2}). 

\noindent \textbf{Discontinuity of the SPP}
Here we show that for the strongly nonreciprocal (unidirectional) case, and for a general nonreciprocal case, near the source point the SPP contribution to the Green function is discontinuous. We also show that for nonreciprocal systems, $\Gamma_{21}>\Gamma_{11}$ can occur. 

\begin{figure}[h]
	\begin{center}
		\noindent
		\includegraphics[width=3.5in]{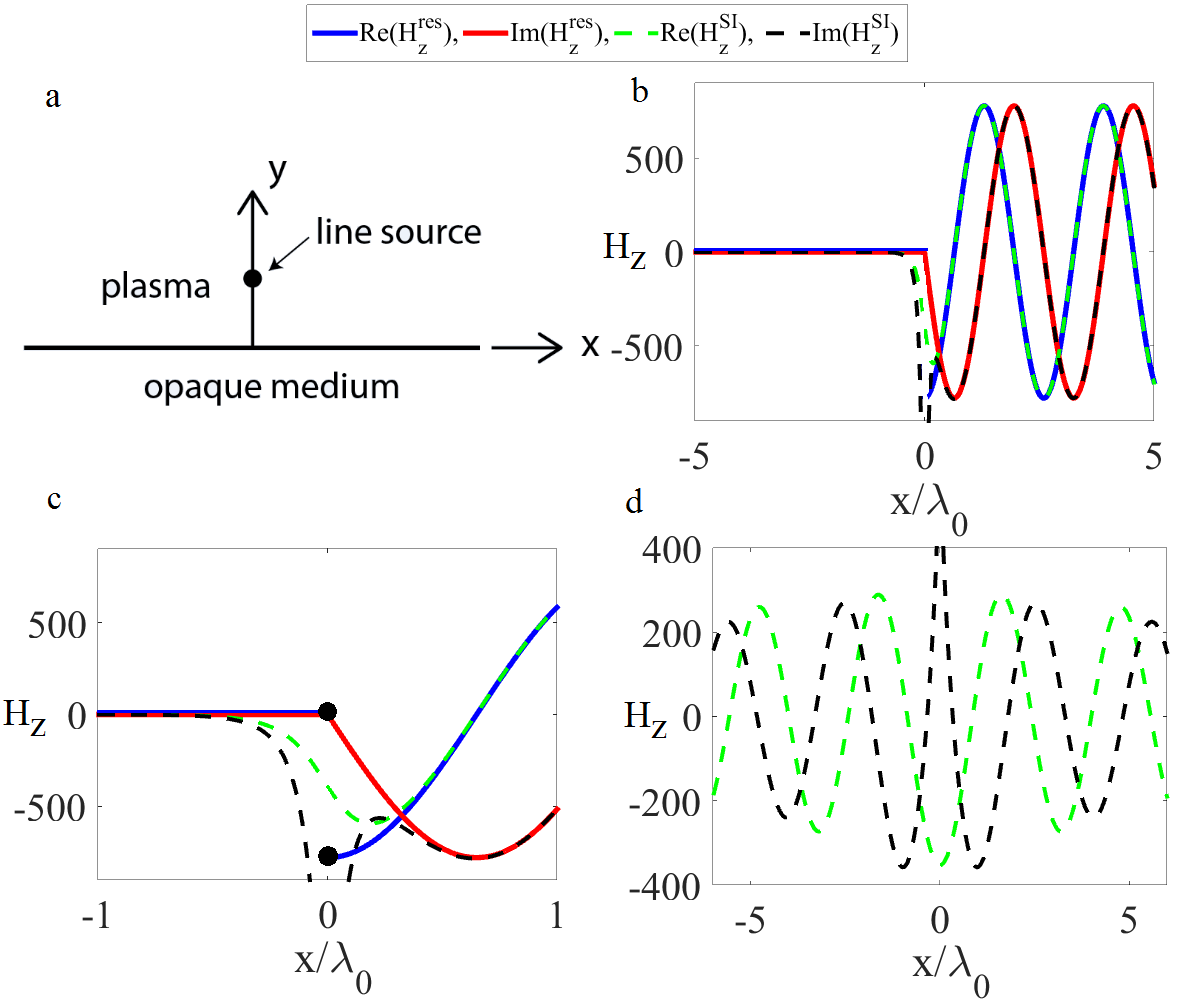}
		\caption{a. Magnetic current source (black dot, $z$-directed and $z$-invariant) located at $x=0, y=d$ inside a biased plasma region, with an opaque half space occupying $y<0$. b. Magnetic field $ \mathrm{H}_z(x) $ at the interface of an $\varepsilon=-2$ half-space and a magnetized plasma having $ \omega_p/\omega = 0.95 $ and $ \omega_c / \omega = 0.21 $, at $\omega_0/2\pi= 200 $ THz. The magnetic line source is located $ \lambda_0/10 $ above the interface in the plasma region, and the field is evaluated at ($x,y=\lambda_0/10,z=0$). c. Field behavior in the vicinity of the source showing the discontinuity of the residue component. d. Same as (b) for the unbiased (reciprocal) case, $ \omega_c / \omega = 0 $.}\label{Hzz}
	\end{center}
\end{figure}

To avoid analytical complications of the general 3D case, we first assume a simple 2D model of a $z$-directed and $z$-invariant magnetic current source located at $x=0, y=d$ inside a biased plasma half-space, adjacent to an opaque half-space occupying $y<0$, as depicted in Fig. \ref{Hzz}a. The resulting magnetic field in the plasma is \cite{bulkedge, SR}
\begin{align}\label{SI}
\mathrm{H}_{z} = \mathrm{H}^{\mathrm{inc}}_z + \frac{\mathrm{A}_0}{2 \pi} \int_{- \infty}^{+ \infty} \frac{1}{2 \gamma_p} \mathrm{R_0}\, e^{-\gamma_p(y+d) + i\mathrm{k}_x x}d\mathrm{k}_x,
\end{align} 
where $ \mathrm{A}_0 = i\omega \varepsilon_0 \varepsilon_{\text{eff}} \mathrm{I}_m $, with $ \mathrm{I}_m $ the magnetic current (set to unity) and $ \mathrm{R_0} $ accounts for the interface, 
\begin{equation}
\mathrm{R_{0}}=\frac{\frac{\gamma _{p}}{\varepsilon _{\text{eff}}}+\frac{%
i\varepsilon _{12}}{\varepsilon _{11}}\frac{i\mathrm{k}_{x}}{\varepsilon _{%
\text{eff}}}-\frac{\gamma _{m}}{\varepsilon _{m}}}{\frac{\gamma _{p}}{%
\varepsilon _{\text{eff}}}-\frac{i\varepsilon _{12}}{\varepsilon _{11}}\frac{%
i\mathrm{k}_{x}}{\varepsilon _{\text{eff}}}+\frac{\gamma _{m}}{\varepsilon
_{m}}},
\end{equation}
where $ \gamma_p = \sqrt{\mathrm{k}_x^2 - \varepsilon_{\text{eff}} \mathrm{k}_0^2  } $, $\gamma_m = \sqrt{\mathrm{k}_x^2 - \varepsilon_{m} \mathrm{k}_0^2  } $ and $ \varepsilon_m $ is the permittivity of the metal (opaque medium). The field in the absence of the interface is
\begin{align}
\mathrm{H}^{\mathrm{inc}}_z &= \frac{\mathrm{A}_0}{2\pi} \int_{-\infty}^{+\infty}  \frac{1}{2 \gamma_p} e^{ -\gamma_p |y-d| + i \mathrm{k}_x x }  d\mathrm{k}_x \\ \notag &= \frac{\textrm{A}_0}{-4i} \mathrm{H}_0^{(1)} \left( \mathrm{k}_0 \sqrt{\varepsilon_{\text{eff}}} \rho  \right)
\end{align}
where $ \mathrm{H}_0^{(1)} $ is the Hankel function of the first kind and order zero and $ \rho = \sqrt{  x^2 + (y-d)^2 } $. The source-point singularity is contained in $\text{Im}(\mathrm{H}_0^{(1)})$, and $\Gamma \sim \text{Im}(\textrm{G}_{yy}) \sim \text{Re}(\textrm{E}_y) \sim \text{Re}(\textrm{H}_z)$.

The interface reflection coefficient $\mathrm{R}_0$, leading to the scattered field, contains pole singularities at the SPP wavenumbers (e.g., the denominator of $\textrm{R}_0$ is the SPP dispersion equation). For $ |\varepsilon_m| \rightarrow \infty $ (perfect conductor), there is one pole at $ k_{\mathrm{spp}, x}  = \pm k_0 \sqrt{\varepsilon_{11}} $ for $\omega_c \gtrless 0$. For $ |\varepsilon_m|$ finite the dispersion equation must be solved numerically, and the plasma may be strongly nonreciprocal, supporting a unidirectional SPP (operating in the bulk bandgap), nonreciprocal, supporting SPPs traveling in opposite directions with unequal wavenumbers (operating above the bulk bandgap), or, in the unbiased (no bandgap) case, reciprocal.

Complex-plane analysis of the magnetic field leads to its evaluation as the sum of a branch cut integral (continuous spectrum) and a discrete residue (SPP) contribution, the latter being 
\begin{align}\label{Eq:Hz_res}
\mathrm{H}_{z}^{\mathrm{res}}& =\theta \left( -x\right) i\mathrm{A}_{0}\text{%
\textrm{Res}}^{\left( -\right) } \frac{%
e^{-\gamma _{p}^{\left( -\right) }(y+d)+i\mathrm{k}_{x,\text{SPP}}^{\left(
-\right) }x}}{2\gamma _{p}^{\left( -\right) }} \\ \notag
& +\theta \left( x\right) i\mathrm{A}_{0}\text{\textrm{Res}}^{\left(
+\right) } \frac{e^{-\gamma _{p}^{\left(
+\right) }(y+d)+i\mathrm{k}_{x,\text{SPP}}^{\left( +\right) }x}}{2\gamma
_{p}^{\left( +\right) }}
\end{align}%
where \textrm{Res}$^{\left( \pm \right) }$ is the residue of $\mathrm{R}_{0}$
evaluated at $\mathrm{k}_{x}=\mathrm{k}_{x,\text{SPP}}^{\left( \pm \right) }$%
,  and $\gamma _{p}^{\left( \pm \right) }=\sqrt{\left( \mathrm{k}_{x,\text{%
SPP}}^{\left( \pm \right) }\right) ^{2}-\epsilon _{eff}\mathrm{k}_{0}^{2}}$,
where $\mathrm{k}_{x,\text{SPP}}^{\left( \pm \right) }$ is the SPP pole for $%
\mathrm{k}_{x}\gtrless 0$ (forward propagating or backward propagating), and where $ \theta(x) $ is the Heaviside step function. In the strongly nonreciprocal (unidirectional) case, only one pole is present, leading to only one term in \eqref{Eq:Hz_res}.

Figure \ref{Hzz}b shows the magnetic field in the bulk bandgap for $\omega_c>0$ obtained by numerical evaluation of the Sommerfeld integral (\ref{SI}), and by assuming only the residue component (\ref{Eq:Hz_res}) (since we operate in the bulk bandgap and the gap Chern number is $-1$, then there is one unidirectional SPP). The opaque medium is topologically-trivial, and is an unbiased plasma having $ \varepsilon = -2 $. As shown in the close-up Fig. \ref{Hzz}c, the residue accurately approximates the field except very close to the source, where the real-part of the residue ($\propto \Gamma_{\textrm{SPP}}$) has an unphysical discontinuity, indicated by the two black dots. In this case, the radiation continuum compensates for the discontinuity of the residue, such that the real-part of the full Sommerfeld integral ($\propto \Gamma$), is continuous, and the SPP peak is pushed away from the source point. 

As a result of the importance of the radiation continuum near the source, at some points $ \mathrm{H}_z(x=0) < \mathrm{H}_z(x > 0) $, so that $ \Gamma_{21} $ exceeds $ \Gamma_{11} $. Figure \ref{Hzz}d shows the unbiased (reciprocal) case for the full Sommerfeld integral, where the field peak occurs at $x=0$ and $\Gamma_{21}<\Gamma_{11}$ at all points. In general, there is a quadrature relationship between the dissipative and coherent rates.

\begin{figure}[h]
	\begin{center}
		\noindent
		\includegraphics[width=3.5in]{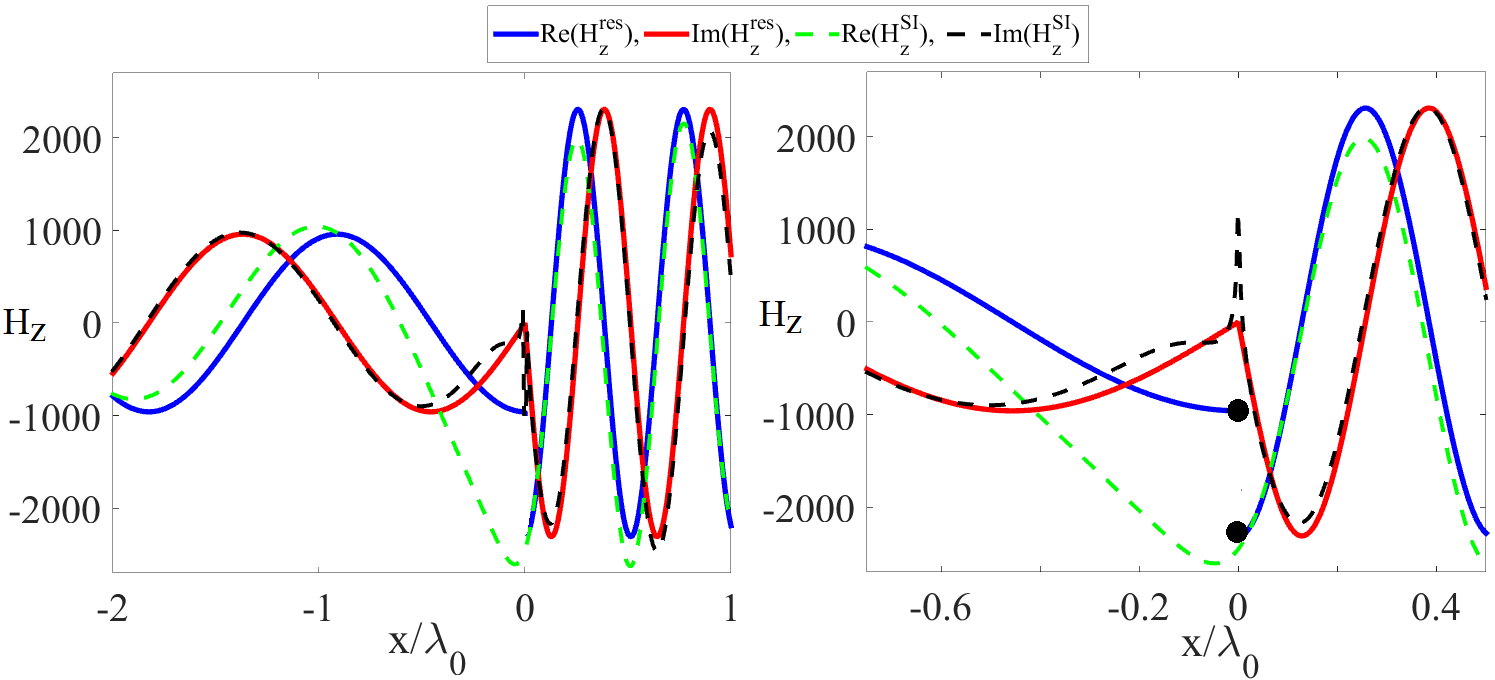}
		\caption{a. Magnetic field $ \mathrm{H}_z(x) $ at the interface of an $\varepsilon=-0.47$ half-space and a magnetized plasma having $ \omega_p/\omega = 0.95 $ and $ \omega_c / \omega = 0.20 $, at $\omega_0/2\pi= 230 $ THz. The magnetic line source is located $ \lambda_0/10 $ above the interface in the plasma region, and the field is evaluated at ($x,y=\lambda_0/10,z=0$).  b. Field behavior in the vicinity of the source showing the discontinuity of the residue component.}\label{Hz2}
	\end{center}
\end{figure}

Figure \ref{Hz2} shows the magnetic field at a frequency outside the bandgap, where we have two SPPs propagating in opposite directions with unequal wavenumbers. As with the unidirectional case, the residue shows a discontinuity at the source point.

\begin{figure}[h]
	\begin{center}
		\noindent
		\includegraphics[width=2.8in]{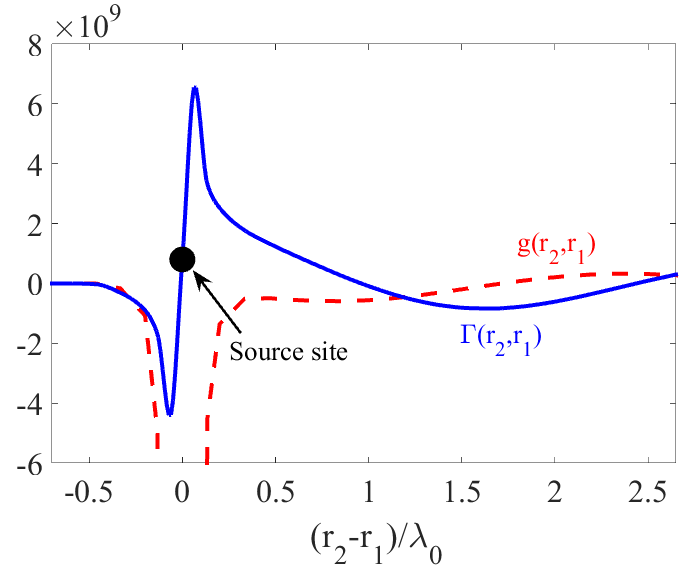}
		\caption{Dissipative decay (solid blue) and coherent (dashed red) rates at the interface of a biased plasma ($ \omega_p/\omega = 0.95 $, $ \omega_c/\omega = 0.21 $) and an opaque medium ($ \varepsilon = -2 $) at $ \omega/ 2 \pi = 200 $ THz. The black circle demonstrate the point dipole source, and the dipole moment is $ \mathrm{d} = 60 $ D.}\label{Gamma_ij}
	\end{center}
\end{figure}

Considering now the 3D case of an electric dipole source at the interface, Fig. \ref{Gamma_ij} shows the dissipative decay and coherent rates (\ref{coupling_terms}) along the interface, computed using the finite element method (COMSOL, \cite{COMSOL}). In this case, it is impossible to separate the discrete and continuum contributions to the field. It can be seen that, as predicted by the previous analytical 2D model, it occurs that $\Gamma$ is nearly discontinuous at the source point (the discontinuity of the discrete spectrum is softened by the radiation continuum), and that $ \Gamma_{21} > \Gamma_{11} $ at some points. The coherent rate becomes unbounded at the source due to the well-known divergence of the real part of the Green function. 

\section*{Appendix III: Concurrence in the unidirectional case}\label{AIII}
In this section we derive the concurrence for a unidirectional system. 

Suppose that the system of qubits are communicating through a strongly nonreciprocal environment, so that the communication is strictly unidirectional, such as occurs for SPPs at PTI interfaces. Assuming that $ {\boldsymbol{\mathrm{G}}}({\boldsymbol{\mathrm{r}}}_1,{\boldsymbol{\mathrm{r}}}_2) $ and $ {\boldsymbol{\mathrm{G}}}({\boldsymbol{\mathrm{r}}}_2,{\boldsymbol{\mathrm{r}}}_1) $ are the dyadic Green function propagators along two opposite directions, the unidirectionality assumption leads to, e.g., $ {\boldsymbol{\mathrm{G}}}({\boldsymbol{\mathrm{r}}}_1,{\boldsymbol{\mathrm{r}}}_2) = 0 $ ($ \Gamma_{12}=g_{12}=0 $) and $ {\boldsymbol{\mathrm{G}}}({\boldsymbol{\mathrm{r}}}_2,{\boldsymbol{\mathrm{r}}}_1) \neq 0 $. 

Under this unidirectionality assumption, the 3D Lindblad superoperator (\ref{Eq:Lindblad_bi}) reduces to
\begin{align}\label{Eq:ME_uni_general}
	\frac{\partial {\rho_s}(t)}{\partial t} &  = -\frac{i}{\hbar} \left[ {\mathrm{H}}_s + \mathrm{V}^{AF}, {\rho_s}(t) \right] \notag \\&
	+ \frac{\Gamma_{11}}{2} \left(  2 {\sigma}_1 {\rho_s}(t) {\sigma}^{\dagger}_1 - {\sigma}^{\dagger}_1 {\sigma}_1 {\rho_s}(t) - {\rho_s}(t)  {\sigma}^{\dagger}_1 {\sigma}_1  \right) \notag \\&   + \frac{\Gamma_{11}}{2}  \left(  2 {\sigma}_2 {\rho_s}(t) {\sigma}^{\dagger}_2 -  {\sigma}^{\dagger}_2 {\sigma}_2 {\rho_s}(t) - {\rho_s}(t)  {\sigma}^{\dagger}_2 {\sigma}_2 \right) \notag \\&
	+ (\frac{\Gamma_{21}}{2} + ig_{21}) \left(   {\sigma}_2 {\rho_s}(t) {\sigma}^{\dagger}_1  - {\rho_s}(t) {\sigma}^{\dagger}_1 {\sigma}_2   \right) \notag \\&
	+ (\frac{\Gamma_{21}}{2} - ig_{21}) \left(   {\sigma}_1 {\rho_s}(t) {\sigma}^{\dagger}_2  -  {\sigma}^{\dagger}_2 {\sigma}_1 {\rho_s}(t)  \right) 
\end{align}
where it has been assumed that $ \Gamma_{11} = \Gamma_{22} $. 

Defining the basis
\begin{align}
&\left|1\right> = \left|g_1\right> \otimes \left|g_2\right> = \left|g_1,g_2\right>,~ \left|2\right> = \left|e_1\right> \otimes \left|e_2\right> = \left|e_1,e_2\right> \notag \\ & \left|3\right> = \left|g_1\right> \otimes \left|e_2\right> = \left|g_1,e_2\right>,~ \left|4\right> = \left|e_1\right> \otimes \left|g_2\right> = \left|e_1,g_2\right>
\end{align}
and considering the system of qubits to be initially prepared in the state $\left|4\right> = \left|e_1\right> \otimes \left|g_2\right>$, it can be shown that for the non-pumped case the non-zero components of the density matrix in (\ref{Eq:ME_uni_general}) are ($\rho$=$\rho_s$)
\begin{align}
& \partial_{t} {\rho}_{11} = \Gamma_{11} (\rho_{33} + \rho_{44}) + \gamma \rho_{34} + \gamma^* \rho_{43} \notag \\&
\partial_t \rho_{33} = - \Gamma_{11} \rho_{33}  - \gamma \rho_{34} - \gamma^* \rho_{43} \notag \\ &
\partial_t \rho_{34} = - \Gamma_{11} \rho_{34} - \gamma^* \rho_{44} \notag \\&
\partial_t \rho_{43} = - \Gamma_{11} \rho_{43} - \gamma \rho_{44} \notag \\&
\partial_t \rho_{44} = -\Gamma_{11} \rho_{44}
\end{align}
where $ \gamma= \Gamma_{21}/2 + ig_{21} $. For all times the density matrix is block diagonal. Concurrence for arbitrary materials can be calculated as \cite{Wootters} 
\begin{equation}\label{concgen}
\mathcal{C} = \mathrm{max}(0,\sqrt{u_1} - \sqrt{u_2} - \sqrt{u_3} - \sqrt{u_4}),
\end{equation}
where $u_i$ are arranged in descending order of the eigenvalues of the matrix $ {\rho}(t) {\rho}^y(t) $, where $ {\rho}^y(t) = \sigma_y \otimes \sigma_y {\rho}^{\star}(t) \sigma_y \otimes \sigma_y$ is the spin-flip density matrix with $\sigma_y$ being the Pauli matrix. We have
\begin{equation}
{\rho}(t) {\rho}^y(t)=
\begin{bmatrix}
0 & 0 & 0 & 0  \\
0 & 0 & 0 & 0  \\
0 & 0 & x & y \\
0 & 0 & z & x
\end{bmatrix}~\rightarrow~ \begin{array}{c}
u_1 = x+\sqrt{yz} \\
u_2 = x-\sqrt{yz}   \\
u_3 = 0  \\
u_4 = 0
\end{array} 
\end{equation}
such that $ x = |\rho_{34}|^2 + \rho_{33}\rho_{44} $, $ y= 2\rho_{34}\rho_{33}$ and $ z = 2\rho_{43}\rho_{44}  $ and
\begin{align}\label{Eq:rho(t)}
& \begin{array}{c}
\rho_{44}(t) = e^{-\Gamma_{11}t} \\
\rho_{43}(t) = - \gamma t e^{-\Gamma_{11}t}\\
\rho_{34}(t) = - \gamma^* t e^{-\Gamma_{11}t} \\
\rho_{33}(t) = |\gamma|^2 t^2 e^{-\Gamma_{11}t} \\
\rho_{11}(t) = 1 - e^{-\Gamma_{11}t} -  |\gamma|^2 t^2 e^{-\Gamma_{11}t},
\end{array}
\end{align}
which leads to \eqref{Eq:Ct}.

%%%%%%%%%%%%%%%%%%%%%%%%%%%%%%%%%%%%%%%%%%%%%%
%%%%%%%%%%%%%%%%%%%%%%%%%%%%%%%%%%%%%%%%%%%%%%

%%%%%%%%%%%%%%%%%%%%%%%%%%%%%%%%%%%%%%%%%%%%%%
%%%%%%%%%%%%%%%%%%%%%%%%%%%%%%%%%%%%%%%%%%%%%%
%%%%%%%%%%%%%%%%%%%%%%%%%%%%%%%%%%%%%%%%%%%%%%

\end{document}